\documentclass[12pt]{article}
\usepackage{latexsym}
\usepackage{amsmath,amsfonts}
\usepackage{times}
\allowdisplaybreaks[4]
\catcode`\@=11

\newcount\hour
\newcount\minute
\newtoks\amorpm \hour=\time\divide\hour by 60\minute
=\time{\multiply\hour by 60 \global\advance\minute by-\hour}
\edef\standardtime{{\ifnum\hour<12 \global\amorpm={am}%
        \else\global\amorpm={pm}\advance\hour by-12 \fi
        \ifnum\hour=0 \hour=12 \fi
        \number\hour:\ifnum\minute<10
        0\fi\number\minute\the\amorpm}}
\edef\militarytime{\number\hour:\ifnum\minute<10
0\fi\number\minute}

\def\draftlabel#1{{\@bsphack\if@filesw {\let\thepage\relax
   \xdef\@gtempa{\write\@auxout{\string
      \newlabel{#1}{{\@currentlabel}{\thepage}}}}}\@gtempa
   \if@nobreak \ifvmode\nobreak\fi\fi\fi\@esphack}
        \gdef\@eqnlabel{#1}}
\def\@eqnlabel{}
\def\@vacuum{}
\def\marginnote#1{}
\def\draftmarginnote#1{\marginpar{\raggedright\scriptsize\tt#1}}
\def\draft{
        \pagestyle{plain}
        \overfullrule=2pt
        \oddsidemargin -.5truein
        \def\@oddhead{\sl \phantom{\today\quad\militarytime} \hfil
        \smash{\Large\sl DRAFT} \hfil \today\quad\militarytime}
        \let\@evenhead\@oddhead
        \let\label=\draftlabel
        \let\marginnote=\draftmarginnote
        \def\ps@empty{\let\@mkboth\@gobbletwo
        \def\@oddfoot{\hfil \smash{\Large\sl DRAFT} \hfil}
        \let\@evenfoot\@oddhead}
        \def\@eqnnum{(\theequation)\rlap{\kern\marginparsep\tt\@eqnlabel}%
        \global\let\@eqnlabel\@vacuum}  }

\newcommand{\rf}[1]{(\ref{#1})}
\renewcommand{\theequation}{\thesection.\arabic{equation}}
\renewcommand{\thefootnote}{\fnsymbol{footnote}}
\newcommand{\newsection}{    % Numeration of eqs. is automatic
\setcounter{equation}{0}\section}
\def\appendix#1{\addtocounter{section}{1}\setcounter{equation}{0}
\renewcommand{\thesection}{\Alph{section}}
\section*{Appendix \thesection\protect\indent \parbox[t]{11.15cm}{#1}}
\addcontentsline{toc}{section}{Appendix \thesection\ \ \ #1}}

\def\EE{{\cal E}}
\def\HH{{\cal H}}
\def\LL{{\cal L}}

\def\half{\frac{1}{2}}

\def\AdSsm{{\scriptscriptstyle \rm AdS}}

\def\be{\begin{equation}}
\def\ee{\end{equation}}
\def\beq{\begin{eqnarray}}
\def\eeq{\end{eqnarray}}

\def\phik{|\phi\rangle}
\def\phibr{\langle\phi|}

\def\alphab{\bar{\alpha}}
\def\upsilonb{\bar{\upsilon}}

\def\gb{\bar{g}}

\def\irm{{\rm i}}

\newcommand{\No}{\mathbb{N}}

\def\ibf{{\bf i}}
\def\iibf{{\bf ii}}
\def\iiibf{{\bf iii}}
\def\ivbf{{\bf iv}}
\def\vbf{{\bf v}}
\def\vibf{{\bf vi}}

\begin{document}

%\draft

\begin{flushright}
FIAN-TD-2017-27 \phantom{\hspace{0.7cm}}   \\
1711.11007 V2 [hep-th]
\end{flushright}

\vspace{1cm}

\begin{center}

{\Large \bf Continuous-spin mixed-symmetry fields in AdS(5)}

\vspace{2.5cm}

R.R. Metsaev\footnote{ E-mail: metsaev@lpi.ru }

\vspace{1cm}

{\it Department of Theoretical Physics, P.N. Lebedev Physical
Institute, \\ Leninsky prospect 53,  Moscow 119991, Russia }

\vspace{3.5cm}

{\bf Abstract}

\end{center}

Free mixed-symmetry continuous-spin fields propagating in AdS(5) space and flat R(4,1) space are studied. In the framework of a light-cone gauge formulation of relativistic dynamics, we build simple actions for such fields. Realization of relativistic symmetries on space of light-cone gauge mixed-symmetry continuous-spin fields is also found. Interrelations between constant parameters entering the light-cone gauge actions and eigenvalues of the Casimir operators of space-time symmetry algebras are obtained. Using these interrelations and requiring that the field dynamics in AdS(5) be irreducible and classically unitary, we derive restrictions on the constant parameters and eigenvalues of the 2nd-order Casimir operator of the so(4,2) algebra.

\newpage
\renewcommand{\thefootnote}{\arabic{footnote}}
\setcounter{footnote}{0}

%%%%%%%%%%%%%%%%%%%%%%%%%%%%%%%%%%%%%%%%%%%%%%%%%%%%%%%%%%%%%%%%%%%%%%%%%%%%%%%%%%%%%%%%%%%%
\section{ \large Introduction}
%%%%%%%%%%%%%%%%%%%%%%%%%%%%%%%%%%%%%%%%%%%%%%%%%%%%%%%%%%%%%%%%%%%%%%%%%%%%%%%%%%%%%%%%%%%%

In view of the aesthetic features of the continuous-spin field theory
a interest in this theory was periodically renewed (see Refs.\cite{Bekaert:2005in}-\cite{Khabarov:2017lth}). Review of recent developments in this topic may be found in Ref.\cite{Bekaert:2017khg}. One of the interesting feature of continuous-spin field is that this field is decomposed into an infinite chain of coupled scalar, vector, and totally symmetric tensor fields which consists of every spin just once. It is such chain of scalar, vector and totally symmetric fields that enters the theory of higher-spin gauge field in AdS space \cite{Vasiliev:1990en}. We think therefore  that certain interesting interrelations between the continuous-spin field theory and the higher-spin gauge theory might exist. Also, as noted in the literature, some regimes in string theory are related to the continuous-spin field theory (see, e.g., Refs.\cite{Savvidy:2003fx}). Regarding string theory, we note that one of the interesting examples of string model is realized as the type IIB superstring in $AdS_5 \times S^5$ background \cite{Metsaev:1998it}. It has been demonstrated in Refs.\cite{Metsaev:2000yf,Metsaev:2000yu} that it the use of a light-cone gauge formulation that considerably simplifies the action of superstring in $AdS_5 \times S^5$ background. Taking this into account it seems then worthwhile to apply a light-cone gauge formulation for studying a continuous-spin field in $AdS_5$ space. This is what we do in this paper.

In this paper, using a light-cone gauge formulation, we study a free mixed-symmetry conti\-nuous-spin bosonic field in $AdS_5$ space. The light-cone gauge formulation of relativistic dynamics in AdS space was developed first in Ref.\cite{Metsaev:1999ui}, while, in Ref.\cite{Metsaev:2002vr}, we applied this formulation for the studying a finite-component mixed-symmetry massless fields in $AdS_5$. Later on, the approach in Ref.\cite{Metsaev:1999ui} has been reformulated into
more convenient form in Ref.\cite{Metsaev:2003cu}, while, in Ref.\cite{Metsaev:2004ee}, we used such renewed light-cone gauge formulation for the studying a finite-component mixed-symmetry massive field in $AdS_5$ space. It is the light-cone gauge formulation in Ref.\cite{Metsaev:2003cu} that we are going to use for the studying a mixed-symmetry continuous-spin field in $AdS_5$ space in this paper.%
\footnote{ Discussion of other approaches which could be helpful for the studying various aspects of mixed-symmetry continuous-spin fields may found in Refs.\cite{Alkalaev:2005kw}-\cite{Alkalaev:2006hq}.
}

We recall that, in manifestly Lorentz covariant formulations, mixed-symmetry fields propagating in 5-dimensional space-time are described by tensor fields whose $so(4,1)$ space-time tensor indices have the structure of the Young tableaux with two rows.%
\footnote{ Recently, in Ref.\cite{Khabarov:2017lth}, two-rows tensor (tensor-spinor)  fields has been used for the frame-like formulation of mixed-symmetry continuous-spin bosonic (fermionic) fields in $AdS_{d+1}$.}
Remarkable feature of the light-cone gauge formulation is that, in the framework of this formulation, mixed-symmetry fields propagating in 5-dimensional space-time are described by complex-valued totally symmetric tensor fields of the $so(3)$ algebra. It is the
use of the complex-valued totally symmetric tensor fields of the $so(3)$ algebra that allows us to obtain the simple light-cone gauge Lagrangian formulation for the mixed-symmetry continuous-spin field in $AdS_5$ space.%
\footnote{ Light-cone gauge approach turns also to be efficient for the studying interacting fields (see, e.g., Refs.\cite{Metsaev:2005ar,Metsaev:2007rn} and references therein). Recent developments of light-cone approach may be found in Refs.\cite{Ponomarev:2016lrm}-\cite{Metsaev:2015rda}.
}
Besides this, our approach allows us in a straightforward way to find interesting restrictions on the eigenvalues of Casimir operators of the $so(4,2)$ algebra which is algebra of relativistic symmetries of continuous-spin field in $AdS_5$. We believe that, in future studies, these restrictions will be helpful for the problem of group theoretical interpretation of continuous-spin field in AdS space.
As by product, we also obtain the light-cone gauge Lagrangian formulation for a mixed-symmetry continuous-spin field in $R^{4,1}$ space.

This  paper is organized as follows.

In Sec.\ref{sec-02},
we discuss a mixed-symmetry continuous-spin field in flat
$R^{4,1}$ space. Namely, for such field, we find a light-cone gauge action and
obtain a light-cone gauge realization of the Poincar\'e algebra symmetries
on the mixed-symmetry continuous-spin field. Also we find interrelations between
constant parameters entering the light-cone gauge formulation of continuous-spin field
and three independent Casimir operators of the Poincar\'e algebra $iso(4,1)$.

In Sec.\ref{sec-03}, we start with a brief review of the general light-cone gauge formalism developed in Ref.\cite{Metsaev:2003cu}. After this, we apply this formalism for the studying mixed-symmetry
continuous-spin field in $AdS_5$ space. Also we present our new result for the light-cone gauge realization of the 3rd-order and 4th-order Casimir operators of the $so(4,2)$ algebra and find interrelations between three constant parameters entering the light-cone gauge Lagrangian formulation of continuous-spin field in $AdS_5$ and three independent Casimir operators of the $so(4,2)$ algebra.
We demonstrate how, for large radius of AdS space, light-cone gauge formulations in AdS space and flat space are related to each other.

In Sec.\ref{sec-04}, we study restrictions imposed on the constant parameters and the 2nd-order Casmir operators of the $so(4,2)$ algebra which are obtained by requiring that the field dynamics
in $AdS_5$ space be classically unitary and irreducible. We find interesting representations
for eigenvalues of the Casimir operators for the mixed-symmetry continuous-spin field which
are similar to the expressions for eigenvalues of the Casimir operators for positive-energy lowest weight unitary representations of the $so(4,2)$ algebra.

In Appendix A, we briefly review the Casimir operators of the $so(4,2)$ algebra. In Appendix B, we present useful relations for various spin operators.

%%%%%%%%%%%%%%%%%%%%%%%%%%%%%%%%%%%%%%%%%%%%%%%%%%%%%%%%%%%%%%%%%%%%%%%%%%%%%%%%
%%%%%%%%%%%%%%%%%%%%%%%%%%%%%%%%%%%%%%%%%%%%%%%%%%%%%%%%%%%%%%%%%%%%%%%%%%%%%%%%
\newsection{ \large Continuous-spin mixed-symmetry field in $R^{4,1}$ space }\label{sec-02}
%%%%%%%%%%%%%%%%%%%%%%%%%%%%%%%%%%%%%%%%%%%%%%%%%%%%%%%%%%%%%%%%%%%%%%%%%%%%%%%%
%%%%%%%%%%%%%%%%%%%%%%%%%%%%%%%%%%%%%%%%%%%%%%%%%%%%%%%%%%%%%%%%%%%%%%%%%%%%%%%%

{\bf Notation and conventions}. Relativistic symmetries of field dynamics in $R^{4,1}$ space are described by the Poincar\'e algebra $iso(4,1)$. We use the following commutation relations for generators of the Poincar\'e algebra $iso(4,1)$:
\be  \label{man-26112017-01}
[P^a,J^{bc}] =\eta^{ab}P^c -\eta^{ac}P^b\,, \qquad [J^{ab},J^{ce}] =\eta^{bc}J^{ae} + 3 \hbox{ terms},
\ee
where $\eta^{ab}$ is a mostly positive flat metric tensor. In this section, vector indices of the $so(4,1)$ Lorentz algebra  take  values $a,b=0,1,2,3,4$. The generators $P^a$, $J^{ab}$ are assumed to be anti-hermitian.

The light-cone frame coordinates and the vector indices of the $so(3)$ algebra are given by
\be  \label{man-26112017-02}
x^\pm = \frac{1}{ \sqrt{2} } (x^4 \pm x^0)\,, \qquad  x^I\,, \qquad I,J,K=1,2,3\,,
\ee
where the coordinate $x^+$ is treated as an evolution parameter. The $so(4,1)$ Lorentz algebra vector $X^a$ is decomposed as $X^+,X^-,X^I$ and a scalar product of Lorentz algebra vectors $X^a$ and $Y^a$ is represented as
\be \label{man-26112017-03}
\eta_{ab} X^a Y^b = X^+Y^- + X^-Y^+ + X^I Y^I\,.
\ee
Relation \rf{man-26112017-03} implies, that, in the light-cone frame, non vanishing elements of the flat metric $\eta_{ab}$ are given by $\eta_{+-}=1$, $\eta_{-+}=1$, $\eta_{IJ} = \delta_{IJ}$. Thus, for the covariant and contravariant components of vectors, we get the relations $X^+=X_-$, $X^-=X_+$, $X^I=X_I$.
In the light-cone frame, commutators of the Poincar\'e algebra generators  are
obtained from the ones in \rf{man-26112017-01} by using the non-vanishing elements of the $\eta^{ab}$ given by $\eta^{+-}=1$, $\eta^{-+}=1$, $\eta^{IJ}=\delta^{IJ}$.

\noindent {\bf Field content}. To discuss light-cone gauge description of a mixed-symmetry continuous-spin field in $R^{4,1}$, we use the following set of complex-valued fields of the $so(3)$ algebra,
\be \label{man-13112017-01}
\phi^{I_1\ldots I_n}(x)\,, \qquad n = h_2, h_2+1,\ldots , \infty\,,
\ee
where $h_2 \in \No$ is a integer which labels the mixed-symmetry continuous-spin field.%
\footnote{ Throughout this paper, $\No$ stands for $1,2,\ldots,\infty$, while $\No_0$ stands for $0,1,2,\ldots,\infty$.}
In \rf{man-13112017-01}, field with $n=1$ is a vector field of the $so(3)$ algebra, while field with $n\geq 2$ is a totally symmetric rank-$n$ traceless tensor field of the $so(3)$ algebra. Note that, in view of $h_2\in \No$, fields $\phi^{I_1\ldots I_n}$ with $n=0,1,\ldots, h_2-1$ do not enter the field content of the mixed-symmetry continuous-spin field \rf{man-13112017-01}. Also, note that field $\phi^{I_1\ldots I_n}$ with $n=0$ stands for a scalar field of the $so(3)$ algebra.

To streamline the presentation,  we introduce creation operators $\alpha^I$, $\upsilon$ and the respective annihilation operators $\alphab^I$, $\upsilonb$ which we refer to as oscillators in this paper. The oscillators, the hermitian conjugation rule, and the vacuum $|0\rangle$ are defined by the relations
\beq
\label{man-27012018-01} && [\alphab^I,\alpha^J] = \delta^{IJ}\,, \qquad [\upsilonb,\upsilon]=1\,, \qquad \alpha^{I\dagger} = \alphab^I\,, \qquad \upsilon^\dagger = \upsilonb\,,
\\
\label{man-27012018-02} && \alphab^I|0\rangle \,, \hspace{2.4cm} \upsilonb |0\rangle = 0 \,.
\eeq
Using the oscillators $\alpha^I$, $\upsilon$, we collect fields \rf{man-13112017-01} into ket-vector defined by
\be \label{man-13112017-02}
\phik = \sum_{n=h_2}^\infty \frac{\upsilon^n}{n!\sqrt{n!}} \alpha^{I_1} \ldots \alpha^{I_n} \phi^{I_1\ldots I_n}(x)|0\rangle\,.
\ee
Ket-vector \rf{man-13112017-02} satisfies the algebraic constraints
\beq
\label{man-13112017-03} && (N_\alpha - N_\upsilon) \phik = 0\,, \hspace{1cm} N_\alpha \equiv \alpha^I\alphab^I\,, \qquad N_\upsilon \equiv \upsilon \upsilonb \,,
\\
\label{man-13112017-04} && \alphab^2 \phik  = 0\,, \hspace{2.6cm} \alphab^2 \equiv \alphab^I\alphab^I\,.
\eeq
From constraint \rf{man-13112017-03}, we learn that the expansion of the $\phik$ \rf{man-13112017-02} into the oscillators $\alpha^I$ and $\upsilon$ involves only those terms of the expansion whose powers in the $\alpha^I$ are equal to powers in the $\upsilon$.
Constraint \rf{man-13112017-04} tells us that fields $\phi^{I_1\ldots I_n}$ \rf{man-13112017-01} are traceless tensor fields of the $so(3)$ algebra.

\noindent {\bf Light-cone gauge action and its relativistic symmetries}. In terms of ket-vector $\phik$ \rf{man-13112017-02}, light-cone gauge action of mixed-symmetry continuous-spin field takes the form
\beq
\label{man-14112017-01} && S  = \int dx^+ dx^- d^3x\, \LL\,, \hspace{1cm} \LL  =  \phibr \bigl(\Box -m^2) \phik\,,
\\
&& \Box = 2\partial^+\partial^- + \partial^I \partial^I \,, \qquad \quad \partial^+ = \partial/\partial x^-\,, \quad \partial^-=\partial/\partial x^+\,, \quad \partial^I= \partial/\partial x^I \,, \hspace{1cm}
\eeq
where a bra-vector $\phibr$ in \rf{man-14112017-01} is obtained from ket-vector $\phik$ \rf{man-13112017-02}
by using the rule $\phibr = \phik^\dagger$.

We now discuss the Poincar\'e algebra $iso(4,1)$ symmetries of
light-cone gauge action \rf{man-14112017-01}. As is known the choice of
the light-cone gauge spoils the manifest $so(4,1)$ Lorentz algebra symmetries.
Therefore in order to show that the Poincar\'e algebra symmetries are still
present, we should find an explicit realization of the Poincar\'e algebra symmetries on ket-vector $\phik$ \rf{man-13112017-02}.

The representation for the generators of the Poincar\'e algebra in terms  of differential operators  acting on ket-vector $|\phi\rangle$ \rf{man-13112017-02} is given by
\beq
\label{man-14112017-04} && P^I = \partial^I\,, \hspace{4.5cm}  P^+=\partial^+\,,
\\
\label{man-14112017-05} && J^{+-} =  x^+ P^- -  x^- \partial^+\,,\hspace{2cm}
 J^{+I} = x^+ \partial^I - x^I\partial^+\,,
\\
\label{man-14112017-06} && J^{IJ} = x^I\partial^J - x^J \partial^I + M^{IJ}\,,
\\
\label{man-14112017-07} && P^- = -\frac{\partial^I\partial^I - m^2}{2\partial^+}\,, \hspace{2.5cm} J^{-I} = x^-\partial^I - x^I  P^- + M^{-I}\,,\qquad
\eeq
where operators $M^{-I}$, $M^{IJ}$ are defined as
\beq
\label{man-14112017-10} && M^{-I} = M^{IJ}\frac{\partial^J}{\partial^+} + \frac{1}{\partial^+} M^I\,,
\\
 \label{man-14112017-11} && \hspace{1.3cm} M^{IJ} =\alpha^I \alphab^J - \alpha^J \alphab^I\,.
\eeq
In \rf{man-14112017-06}, \rf{man-14112017-10}, a quantity $M^{IJ}$ stands for a spin operator of the $so(3)$ algebra. In \rf{man-14112017-11}, we present the well known realization of the $M^{IJ}$ on ket-vector $\phik$ \rf{man-13112017-02}. Operator $M^I$ \rf{man-14112017-10} does not depend on space-time coordinates and their derivatives. This operator acts only on spin indices of ket-vector $\phik$ \rf{man-13112017-02}. The operator $M^I$ transforms as a vector of the $so(3)$ algebra,
\be \label{man-14112017-12}
[M^I,M^{JK}] = \delta^{IJ} M^K - \delta^{IK} M^J\,,
\ee
and satisfies the following commutation relations
\be \label{man-14112017-14}
[M^I,M^J] = m^2 M^{IJ}\,.
\ee
It is the equations \rf{man-14112017-14} that are the basic equations of the light-cone gauge formulation of relativistic dynamics in the flat space. In the framework of the light-cone gauge formulation, the most difficult problem is to find a solution to the basic equations \rf{man-14112017-14}.

We now discuss our solution for the operator $M^I$ corresponding to the mixed-symmetry continuous-spin field in $R^{4,1}$. Solution for the operator $M^I$ we found is given by
\beq
\label{man-15112017-01} M^I & = &  l S^I + g \alphab^I + A^I \gb\,,
\\
\label{man-15112017-02} && S^I \equiv \epsilon^{IJK}\alpha^J \alphab^K \,,
\\
\label{man-15112017-03} && A^I \equiv \alpha^I -  \alpha^2 \frac{1}{2N_\alpha+3} \alphab^I\,,
\\
\label{man-15112017-04} && N_\alpha \equiv \alpha^I\alphab^I\,, \qquad N_\upsilon \equiv \upsilon \upsilonb \,, \qquad  \alpha^2 \equiv \alpha^I\alpha^I\,,
\\
\label{man-15112017-06} && l \equiv \frac{\irm h_2 \kappa}{N_\upsilon (N_\upsilon +1)}\,,
\\
\label{man-15112017-05} && g \equiv g_\upsilon \upsilonb \,, \qquad \gb \equiv  \upsilon g_\upsilon \,,
\qquad g_\upsilon \equiv \Bigl[\frac{((N_\upsilon+1)^2 - h_2^2)  }{(N_\upsilon+1)^3(2N_\upsilon+3)} F_\upsilon \Bigr]^{1/2}\,,\qquad
\\
\label{man-15112017-08} && F_\upsilon \equiv     \kappa^2 - (N_\upsilon+1)^2 m^2  \,,
\eeq
where $\epsilon^{IJK}$ \rf{man-15112017-02} stands for the Levi-Civita symbol of rank three with $\epsilon^{123}=1$. In \rf{man-15112017-06},\rf{man-15112017-08}, a quantity $\kappa$ stands for a dimensionfull constant parameter.

The following remarks are in order.

\noindent \ibf) From \rf{man-15112017-06}-\rf{man-15112017-08}, we see that the mixed-symmetry continuous-spin field in $R^{4,1}$ space
is labeled by three parameters: one integer $h_2\in \No$, and two dimensionfull parameters $m$ and $\kappa$.

\noindent \iibf) In this paper, field $\phik$ \rf{man-13112017-02} having $m=0$ is referred to as massless continuous-spin field, while field $\phik$ \rf{man-13112017-02} having $m\ne0$ is referred to as massive continuous-spin field.%
\footnote{ For the massless continuous-spin massless propagating in $R^{4,1}$, the  discussion of the operator $M^I$ can also be found in Sec.2 in Ref.\cite{Brink:2002zx}.}

\noindent \iiibf) if $\kappa h_2 = 0$, then, from \rf{man-15112017-06}, we see that operator $M^I$ \rf{man-15112017-01} becomes real-valued. Therefore complex-valued fields \rf{man-13112017-01} can be restricted to be real-valued. This case corresponds to totally symmetric continuous-spin field.

\noindent \ivbf) If $\kappa h_2 \ne 0$, then considering  $lS^I$-term \rf{man-15112017-06} and   requiring the operator $M^I$ \rf{man-15112017-01} to be hermitian, we find that $\kappa h_2$ should be real-valued. This implies that the $\kappa$ should be real-valued. For definiteness, we assume that the $\kappa$ is strictly positive. Thus we have the classification
\beq
\label{man-15112017-08-a1} && \kappa > 0 \,, \qquad \quad h_2 \in \No\,,  \hspace{2.2cm} \hbox{ for mixed-symmetry field};
\\
\label{man-15112017-08-a2} && \kappa h_2 = 0 \,, \qquad h_2 \in \No_0\,,  \hspace{2cm} \hbox{ for totally symmetric field}.
\eeq

\noindent \vbf) Using \rf{man-14112017-04}-\rf{man-14112017-07}, we verify that the light-cone gauge action \rf{man-14112017-01} is invariant under the transformations of the Poincar\'e algebra $iso(4,1)$ algebra given by
\be \label{man-15112017-08-a0}
\delta_G |\phi\rangle = G|\phi\rangle\,,
\ee
where $G$ appearing on r.h.s \rf{man-15112017-08-a0} stands for differential operators given in \rf{man-14112017-04}-\rf{man-14112017-07}.

As is known, the Poincar\'e algebra $iso(4,1)$ has three independent Casimir operators.
Now our aim is to express eigenvalues of the Casimir operators in terms of the three parameters $m$, $\kappa$, $h_2$.

\noindent {\bf Casimir operators of the Poincar\'e algebra $iso(4,1)$}. The three independent Casimir operators, which we denote as $C_2$, $ C_{\epsilon\, 3}$, $C_4$, can be expressed in terms of the generators of the Poincar\'e algebra \rf{man-26112017-01} as
\beq
\label{man-26112017-09} && C_2 = P^a P^a\,,
\\
\label{man-26112017-10} && C_{\epsilon\, 3} = -\frac{\irm }{8}\epsilon^{a_1\ldots a_5} J^{a_1a_2} J^{a_3a_4} P^{a_5} \,,
\\
\label{man-26112017-11} && C_4 = J^{ac} J^{bc} P^a P^b - \half P^c P^c J^{ab}J^{ab}\,,
\eeq
where $\epsilon^{a_1\ldots a_5}$ \rf{man-26112017-10} stands for the Levi-Civita symbol of rank five with $\epsilon^{01234}=1$. Note that operator $C_4$ \rf{man-26112017-11} admits the following representation:
\be
\label{man-26112017-12}  C_4 = \frac{1}{8} C_{\epsilon\,2}^{ab} C_{\epsilon\,2}^{ab} \,, \qquad  C_{\epsilon\,2}^{ab} \equiv  \epsilon^{ab c_1c_2 c_3 } P^{c_1} J^{c_2c_3}\,.
\ee
Plugging the generators of the Poincar\'e algebra \rf{man-14112017-04}-\rf{man-14112017-07} into \rf{man-26112017-09}-\rf{man-26112017-11}, we find that the operator $C_2$ \rf{man-26112017-09} is diagonalized,
\be \label{man-26112017-04}
C_2 = m^2\,,
\ee
while the operators $C_{\epsilon\, 3}$, $C_4$ \rf{man-26112017-10},\rf{man-26112017-11} take the form
\beq
 \label{man-26112017-05} && C_{\epsilon\, 3} = \frac{\irm }{2} \epsilon^{ IJK} M^I M^{JK},
\\
 \label{man-26112017-06} && C_4 = M^I M^I - \half m^2  M^{IJ}M^{IJ}\,.
\eeq
Finally, plugging the operators $M^{IJ}$, $M^I$ \rf{man-14112017-11}, \rf{man-15112017-01} into \rf{man-26112017-05},\rf{man-26112017-06}, we find that the operators $C_{\epsilon\, 3}$, $C_4$ are also diagonalized,
\beq
\label{man-26112017-07} && C_{\epsilon\, 3}  = \kappa h_2\,,
\\
\label{man-26112017-08} && C_4 = \kappa^2 + m^2 (h_2^2-1)\,.
\eeq
From \rf{man-26112017-04},\rf{man-26112017-07},\rf{man-26112017-08}, we see how the eigenvalues of the three Casimir operators $C_2$, $C_{\epsilon\, 3}$, $C_4$ are expressed in terms of the three parameters $m$, $\kappa$, $h_2$.
Note that, eigenvalues of the $C_2$ for the ket-vector $\phik$ and the bra-vector $\phibr$, $\phibr\equiv \phik^\dagger$ are equal. The same holds true for eigenvalues of the $C_4$. Contrary this, eigenvalue of $C_{\epsilon,3}$ \rf{man-26112017-05} for the ket-vector $\phik$  is equal to $\kappa h_2$, while eigenvalue of the $C_{\epsilon,3}$ for the bra-vector $\phibr$  is equal to $-\kappa h_2$.%
\footnote{ First, in the framework of light-cone approach, Casimir operator $C_{\epsilon\, 3}$ \rf{man-26112017-05} and its eigenvalue in \rf{man-26112017-07} were obtained in Ref.\cite{Brink:2002zx}.}

\noindent {\bf Irreducible classically unitary mixed-symmetry continuous-spin field}. Detailed definition of classical unitarity and irreducibility of field dynamics may be found below in Sec.\ref{sec-04}. Briefly speaking, for the mixed-symmetry continuous-spin field in the flat space, the classical unitarity amounts to the two conditions: a) the operator $M^I$ \rf{man-15112017-01} should be hermitian; b) the $F_\upsilon$ \rf{man-15112017-08} should be non-negative, $F_\upsilon\geq 0$, for all $N_\upsilon = h_2,h_2+1,\ldots,\infty$. The irreducibility amounts to the condition $F_\upsilon \ne 0$, for all $N_\upsilon = h_2,h_2+1,\ldots,\infty$. If, for some value of $N_\upsilon=s$, the  $F_\upsilon$ is equal to zero, then the field dynamics is refereed to as reducible field dynamics.

Let us first discuss the irreducible classically unitary mixed-symmetry continuous-spin field. As we noted earlier, the hermicity of operator $M^I$ \rf{man-15112017-01} implies that $\kappa$ should be real-valued. Note also that, for the mixed-symmetry field, $\kappa\ne 0$ \rf{man-15112017-08-a1}. Taking this into account, we see that requiring $F_\upsilon > 0$ for all $N_\upsilon = h_2,h_2+1,\ldots,\infty$, we find the inequality $m^2 \leq 0$. The cases $m^2=0$ and $m^2 \ne 0$ we refer to as massless and massive continuous-spin fields respectively. Thus, we see that the massive continuous-spin field has tachyonic mass. In Ref.\cite{Metsaev:2016lhs}, we conjectured that the massive continuous-spin field is associated with the tachyonic UIR of the Poincar\'e algebra. Discussion of the tachyonic UIR of the Poincar\'e algebra may be found in Ref.\cite{Bekaert:2006py}.

\noindent {\bf Reducible mixed-symmetry continuous-spin field}. Now let us discuss the reducible mixed-symmetry continuous-spin field. Requiring $F_\upsilon|_{N_\upsilon=h_1}=0$, we find the relation
\be \label{man-30012017-01}
\kappa^2 = (h_1+1)^2 m^2\,.
\ee
Plugging $\kappa^2$ \rf{man-30012017-01} into \rf{man-15112017-08}, we get
\be \label{man-30012017-02}
F_\upsilon=(h_1-N_\upsilon)(h_1+2 + N_\upsilon)m^2\,.
\ee
Using $F_\upsilon$ \rf{man-30012017-02}, we can verify that Lagrangian \rf{man-14112017-01} and Poincar\'e algebra transformations \rf{man-15112017-08-a0}  describe the reducible mixed-symmetry continuous-spin field.
Namely, decomposing $\phik$ \rf{man-13112017-02} as
\beq
\label{man-30012017-03} \phik & = &  |\phi^{h_2,h_1}\rangle  + |\phi^{h_1+1,\infty}\rangle \,,
\\
\label{man-30012017-04} && |\phi^{M,N}\rangle \equiv  \sum_{n=M}^N \frac{\upsilon^n}{n!\sqrt{n!}} \alpha^{I_1} \ldots \alpha^{I_n} \phi^{I_1\ldots I_n}(x) |0\rangle\,,
\eeq
we can verify that Lagrangian \rf{man-14112017-01} is factorized as
\beq
\label{man-30012017-05} \LL & = &  \LL^{h_2,h_1} + \LL^{h_1+1,\infty}\,,
\\
\label{man-30012017-06} && \LL^{M_,N}  \equiv   \langle \phi^{M,N} |\bigl(\Box - m^2  \bigr) |\phi^{M,N} \rangle\,.
\eeq
This is to say that $\LL^{h_2,h_1}$ and $\LL^{h_1+1,\infty}$ \rf{man-30012017-05}  are invariant under the Poincar\'e algebra transformations \rf{man-15112017-08-a0}. Using \rf{man-30012017-02} and considering $m^2>0$, we see that $F_\upsilon>0$ when $N_\upsilon = h_2,h_2+1,\ldots, h_1-1$ and $F_\upsilon < 0$ when $N_\upsilon = h_1+1,h_1+2,\ldots, \infty$. This implies that, for $m^2>0$, the $|\phi^{h_2,h_1}\rangle$ \rf{man-30012017-03} describes classically unitary massive finite-component field, while the $|\phi^{h_1+1,\infty}\rangle$ \rf{man-30012017-03} describes classically non-unitary infinite-component field. Note also that, for $m^2<0$, the $\kappa$ becomes imaginary in view of \rf{man-30012017-01}. This implies that, for $m^2<0$, both the $|\phi^{h_2,h_1}\rangle$ and $|\phi^{h_1+1,\infty}\rangle$ \rf{man-30012017-03} describe classically non-unitary mixed-symmetry massive fields.

\noindent {\bf Totally symmetric continuous-spin field}. From \rf{man-15112017-08-a2}, we see that the totally symmetric continuous-spin field is realized by considering the following two cases:
\beq
\label{man-31012017-01}  && h_2  = 0 \,, \hspace{1cm} \kappa -\hbox{arbitrary} \,,
\\
\label{man-31012017-02} && h_2  \ne 0 \,, \hspace{1cm} \kappa = 0 \,.
\eeq

\noindent {\bf Case $h_2=0$, $\kappa$-arbitrary}. Setting $h_2=0$ in \rf{man-13112017-02}, we get ket-vector $\phik$ entering Lagrangian for totally symmetric field \rf{man-14112017-01}.  Also, setting $h_2=0$ in \rf{man-15112017-01}-\rf{man-15112017-08}, we see that the operator $M^I$ is simplified as
\beq
\label{man-31012017-03} && M^I =  g \alphab^I + A^I \gb\,,
\\
\label{man-31012017-04} &&   g = g_\upsilon \upsilonb \,, \qquad \gb =  \upsilon g_\upsilon \,, \qquad  g_\upsilon = \Bigl( \frac{ F_\upsilon }{(N_\upsilon+1) (2N_\upsilon+3)} \Bigr)^{1/2}\,,
\\
\label{man-31012017-05} && F_\upsilon =     \kappa^2 - (N_\upsilon+1)^2 m^2\,,
\eeq
where the operators $A^I$, $N_\upsilon$ takes the same form as in \rf{man-15112017-03},\rf{man-15112017-04}.

The following remarks are in order:

\noindent \ibf) As the $lS^I$-term \rf{man-15112017-01} does not appear in \rf{man-31012017-03},  the hermitian operator $M^I$ \rf{man-31012017-03} turns out to be real-valued. For this reason the complex-valued fields \rf{man-13112017-01} can be restricted to be real-valued.

\noindent \iibf) For the mixed-symmetry field, the hermicity of the operator $M^I$ \rf{man-31012017-03}, in view of the $lS^I$-term \rf{man-31012017-03}, implies that the $\kappa$ should be real-valued. For the totally symmetric field, the $lS^I$-term \rf{man-15112017-01} does not appear in \rf{man-31012017-03}. Therefore, for the totally symmetric field, the hermicity of the operator $M^I$ \rf{man-31012017-03} does not imply that only real-valued $\kappa$ is admitted.

\noindent \iiibf) As the $lS^I$-term \rf{man-15112017-01} does not appear in \rf{man-31012017-03}, all that is required for the classical unitarity and irreducibility of totally symmetric field is to respect the condition $F_\upsilon>0$ for all $N_\upsilon=0,1,\ldots,\infty$. From \rf{man-31012017-05}, we see that the just mentioned condition is satisfied provided
\be \label{man-31012017-06}
\kappa^2 > m^2 \,, \qquad m^2 \leq  0\,.
\ee
From \rf{man-31012017-06}, we see that, for massless field, $m=0$, the parameter $\kappa$ should be real-valued, while, for massive field, $m^2<0$, the parameter  $\kappa$ can be real-valued or purely imaginary.

\noindent \ivbf) To get the reducible totally symmetric field we consider equation $F_\upsilon|_{N_\upsilon=s}=0$. Solution of this equation is given by $\kappa^2 = (s+1)^2m^2$.
Using such solution in \rf{man-31012017-05}, we get
\be \label{man-01022017-01}
F_\upsilon=(s-N_\upsilon)(s+2 + N_\upsilon)m^2\,.
\ee
Now decomposing $\phik$ \rf{man-13112017-02} as
\be \label{man-01022017-02}
\phik = |\phi^{0,s}\rangle  + |\phi^{s+1,\infty}\rangle \,,
\ee
where $|\phi^{0,s}\rangle$, $|\phi^{s+1,\infty}\rangle$ are defined as in \rf{man-30012017-04},  we can verify that Lagrangian \rf{man-14112017-01} is factorized as
\be  \label{man-01022017-03}
\LL  =   \LL^{0,s} + \LL^{s+1,\infty}\,,
\ee
where $\LL^{0,s}$, $\LL^{s+1,\infty}$ are defined as in \rf{man-30012017-06}. This is to say that $\LL^{0,s}$ and $\LL^{s+1,\infty}$ \rf{man-01022017-03}  are invariant under the Poincar\'e algebra transformations \rf{man-15112017-08-a0}. Using \rf{man-01022017-01} and considering $m^2>0$, we verify that $F_\upsilon>0$ when $N_\upsilon = 0,1,\ldots, s-1$ and $F_\upsilon < 0$ when $N_\upsilon = s+1,s+2,\ldots, \infty$. This implies that, for $m^2>0$, the $|\phi^{0,s}\rangle$ in \rf{man-01022017-02} describes classically unitary massive spin-$s$ field, while the $|\phi^{s+1,\infty}\rangle$  in \rf{man-01022017-02} describes classically non-unitary infinite-component field.
On the contrary, for $m^2<0$, we have $F_\upsilon<0$ when $N_\upsilon = 0,1,\ldots, s-1$ and $F_\upsilon > 0$ when $N_\upsilon = s+1,s+2,\ldots, \infty$. Therefore, for $m^2<0$, the $|\phi^{0,s}\rangle$  in \rf{man-01022017-02} describes classically non-unitary massive spin-$s$ field,
while the $|\phi^{s+1,\infty}\rangle$  in \rf{man-01022017-02} describes classically unitary infinite-component field.

\noindent {\bf Case $h_2 \ne 0$, $\kappa=0$}. Totally symmetric field given in \rf{man-31012017-02} turns out to be equivalent to the infinite component field $|\phi^{s+1,\infty}\rangle$ \rf{man-01022017-02}. To see this we set $\kappa=0$ in \rf{man-15112017-01}-\rf{man-15112017-08} and, on the one hand, we obtain the operator $M^I$ as in \rf{man-31012017-03},\rf{man-31012017-04} with the following expression for $g_\upsilon$
\be  \label{man-02022017-01}
g_\upsilon = \Bigl( \frac{ (h_2^2 - (N_\upsilon+1)^2) }{ (N_\upsilon+1) (2N_\upsilon+3)  } m^2 \Bigr)^{1/2}\,.
\ee
On the other hand, the infinite component $|\phi^{s+1,\infty}\rangle$ \rf{man-01022017-02} is described by $g_\upsilon$ given in \rf{man-31012017-04} with $F_\upsilon$ as in \rf{man-01022017-01}\,,
\be  \label{man-02022017-02}
g_\upsilon = \Bigl( \frac{ ((s+1)^2 - (N_\upsilon+1)^2) }{ (N_\upsilon+1) (2N_\upsilon+3)  } m^2 \Bigr)^{1/2}\,.
\ee
Making the identification $h_2 = s + 1 $, we see that expressions for $g_\upsilon$ in \rf{man-02022017-01} and \rf{man-02022017-02} coincide.

%%%%%%%%%%%%%%%%%%%%%%%%%%%%%%%%%%%%%%%%%%%%%%%%%%%%%%%%%%%%%%%%%%%%%%%%%%%%%%%%%
%%%%%%%%%%%%%%%%%%%%%%%%%%%%%%%%%%%%%%%%%%%%%%%%%%%%%%%%%%%%%%%%%%%%%%%%%%%%%%%%%
\newsection{ \large Continuous-spin mixed-symmetry field in $AdS_5$ space } \label{sec-03}
%%%%%%%%%%%%%%%%%%%%%%%%%%%%%%%%%%%%%%%%%%%%%%%%%%%%%%%%%%%%%%%%%%%%%%%%%%%%%%%%%
%%%%%%%%%%%%%%%%%%%%%%%%%%%%%%%%%%%%%%%%%%%%%%%%%%%%%%%%%%%%%%%%%%%%%%%%%%%%%%%%%

{\bf Notation and conventions}. Relativistic symmetries of field dynamics in $AdS_5$ space are described by the $so(4,2)$ algebra. We use the following commutators for generators of the $so(4,2)$ algebra
\beq
&& {}[D,P^a]=-P^a\,, \hspace{2cm}  {}[P^a,J^{bc}]=\eta^{ab}P^c
-\eta^{ac}P^b, \qquad
\nonumber\\
\label{man-27112017-36} && [D,K^a]=K^a\,, \hspace{2.3cm} [K^a,J^{bc}]=\eta^{ab}K^c - \eta^{ac}K^b,
\qquad
\\
&& [P^a,K^b]=\eta^{ab}D - J^{ab}\,, \qquad  [J^{ab},J^{ce}]=\eta^{bc}J^{ae}+3\hbox{ terms} \,,
\nonumber
\eeq
where $\eta^{ab}$ stands for the mostly positive flat metric tensor.  The vector indices of the $so(3,1)$ Lorentz algebra take values $a,b=0,1,2,3$. The generators $P^a$, $K^a$, $D$, $J^{ab}$ are assumed to be anti-hermitian.

We use the Poincar\'e parametrization of $AdS_5$ space,
\be \label{man-27112017-37}
ds^2= \frac{R^2}{z^2}(- dx^0 dx^0 + dx^idx^i + dx^3 dx^3 + dz dz)\,,
\ee
where $R$ is the radius of $AdS_5$ space. The vector indices of the $so(2)$ algebra take values
\be \label{man-27112017-38}
i,j,k =1, 2\,.
\ee
The light-cone frame coordinates $x^\pm$, $x^I$ and their derivatives $\partial^\pm$, $\partial^I$ are defined as
\beq
\label{man-27112017-39} && x^\pm =  \frac{1}{\sqrt{2}} (x^3 \pm x^0)\,, \qquad  x^I = x^i,z\,,
\\
\label{man-27112017-39-a1} && \partial^+=\partial/\partial x^-\,, \quad \partial^-=\partial/\partial x^+\,, \qquad \partial^I=\partial^i,\partial^z\,,\quad  \partial^i=\partial/\partial x^i\,,\quad \partial^z=\partial/\partial z\,,\qquad
\eeq
where $x^+$ is taken to be the light-cone time. In light-cone frame, the $so(3,1)$ Lorentz algebra vector $X^a$ is decomposed as $X^+,X^-,X^i$. A scalar product of the $so(3,1)$ Lorentz algebra vectors $X^a$ and $Y^a$ is represented then as
\be \label{man-27112017-40}
\eta_{ab} X^a Y^b =  X^+Y^- + X^-Y^+ +  X^i Y^i\,.
\ee
From \rf{man-27112017-40}, we see that, in the light-cone frame, non vanishing elements of the flat metric $\eta_{ab}$ are given by $\eta_{+-} =1$, $\eta_{-+}=1$, $\eta_{ij} = \delta_{ij}$, where $\delta_{ij}$ is Kronecker delta symbol. Therefore, in the light-cone frame, commutators for generators of the $so(4,2)$ algebra are obtained from the ones in \rf{man-27112017-36} by using values of the inverse flat metric given by $\eta^{+-}=1$, $\eta^{-+}=1$, $\eta^{ij}=\delta^{ij}$.

In accordance with the decomposition for the coordinates $x^I$ \rf{man-27112017-39}, the $so(3)$ algebra vector $X^I$ is decomposed as $X^i$, $X^z$. A scalar product of the $so(3)$ algebra vectors $X^I=X^i,X^z$ and $Y^I=Y^i,Y^z$ is represented then as
\be
X^I Y^I = X^i Y^i + X^z Y^z\,.
\ee
In what follows a natation $\delta^{IJ}$ stands for Kronecker delta symbol. This symbol is decomposed as $\delta^{IJ} = \delta^{ij},\delta^{zz}$, where $\delta^{zz}=1$.

\noindent {\bf Field content}. To discuss light-cone gauge description of a mixed-symmetry continuous-spin field propagating in $AdS_5$ space we use the following set of {\it complex-valued} fields of the  $so(3)$ algebra
\be  \label{man-14112017-15}
\phi^{I_1\ldots I_n}(x,z)\,, \qquad n = h_2, h_2+1,\ldots , \infty\,,
\ee
where $h_2 \in \No$ is a integer which labels the mixed-symmetry continuous-spin field. In \rf{man-14112017-15}, field with $n=1$ is a vector field of the $so(3)$ algebra, while field with $n\geq 2$ is a totally symmetric rank-$n$ traceless tensor field of the $so(3)$ algebra. Note that, in view of $h_2\in \No$, fields $\phi^{I_1\ldots I_n}$ with $n=0,1,\ldots, h_2-1$ do not enter the field content of the mixed-symmetry continuous-spin field \rf{man-14112017-15}. Also note that field $\phi^{I_1\ldots I_n}$ with $n=0$ stands for a scalar field of the $so(3)$ algebra.

Using the oscillators $\alpha^I$, $\upsilon$ \rf{man-27012018-01}, \rf{man-27012018-02}, we collect fields \rf{man-14112017-15} into ket-vector defined by
\be  \label{man-14112017-16}
\phik = \sum_{n=h_2}^\infty \frac{\upsilon^n}{n!\sqrt{n!}} \alpha^{I_1} \ldots \alpha^{I_n} \phi^{I_1\ldots I_n}(x,z)|0\rangle\,.
\ee
Ket-vector \rf{man-14112017-16} satisfies the algebraic constraints
\beq
\label{man-14112017-17} && (N_\alpha - N_\upsilon) \phik = 0\,, \hspace{1cm} N_\alpha \equiv \alpha^I\alphab^I\,, \qquad N_\upsilon \equiv \upsilon \upsilonb \,,
\\
\label{man-14112017-18} && \alphab^2 \phik  = 0\,, \hspace{2.6cm} \alphab^2 \equiv \alphab^I\alphab^I\,.
\eeq
Constraint \rf{man-14112017-18} amounts to the requirement that fields $\phi^{I_1\ldots I_n}$ \rf{man-14112017-15} are traceless tensor fields of the $so(3)$ algebra.

\noindent {\bf Light-cone gauge action}. In terms of ket-vector $\phik$ \rf{man-14112017-16}, light-cone gauge action of the mixed-symmetry continuous-spin field takes the form
\beq
\label{man-15112017-09} && S  = \int dz d^4x\, \LL\,, \hspace{2.6cm} d^4 x = dx^+ dx^- d^2x\,,
\\
\label{man-15112017-10} && \LL  =   \langle \phi|\bigl(\Box -\frac{1}{z^2}A\bigr)|\phi\rangle\,, \hspace{1.4cm} \Box = 2\partial^+\partial^- + \partial^i \partial^i + \partial^z \partial^z\,,
\eeq
where the D'Alembertian operator \rf{man-15112017-10} takes the same form as in the flat space.
In \rf{man-15112017-10}, a bra-vector $\phibr$ is obtained from ket-vector $\phik$ \rf{man-14112017-16}
by using the rule $\phibr = \phik^\dagger$.
An operator $A$ appearing in \rf{man-15112017-10} does not depend on the space-time coordinates and
their derivatives. This operator acts only on spin indices of the ket-vector $|\phi\rangle$ \rf{man-14112017-16}.
For the reader convenience, we note that for a massive scalar field in $AdS_{d+1}$, the
operator $A$ takes the form
\be \label{man-15112017-12}
A = m^2 R^2 + \frac{d^2-1}{4}\,,
\ee
where $m$ stands for mass parameter of the scalar field.

\noindent {\bf Realization of relativistic symmetries}. As relativistic symmetries of fields in $AdS_5$ space are described by the $so(4,2)$ algebra we now discuss the $so(4,2)$ algebra symmetries of  light-cone gauge action \rf{man-15112017-09}. A choice of the light-cone gauge spoils the manifest $so(3,1)$ Lorentz algebra symmetries. Therefore in order to demonstrate that the symmetries of $so(4,2)$ algebra are still present we should find an explicit realization of the $so(4,2)$ algebra symmetries on the ket-vector $\phik$ \rf{man-14112017-16}. Now, using the general light-cone gauge approach in Ref.\cite{Metsaev:2003cu}, we proceed to discussion of  the light-cone gauge realization of the $so(4,2)$ algebra symmetries on the ket-vector $\phik$ \rf{man-14112017-16}.

The representation for the generators of the $so(4,2)$ algebra in terms  of differential operators acting on the ket-vector $|\phi\rangle$ \rf{man-14112017-16} is given by
\beq
\label{man-15112017-14} && P^i=\partial^i\,, \hspace{3.5cm}  P^+=\partial^+\,,
\\
\label{man-15112017-15} && J^{+-}=x^+ P^- -x^-\partial^+\,, \hspace{1cm} J^{+i}=x^+\partial^i-x^i\partial^+\,,
\\
\label{man-15112017-16} && J^{ij} = x^i\partial^j-x^j\partial^i + M^{ij}\,,
\\
\label{man-15112017-17} && D=x^+ P^- +x^-\partial^++x^I\partial^I + \frac{3}{2}\,,
\\
\label{man-15112017-18} && K^+ = -\frac{1}{2}(2x^+x^-+x^Jx^J)\partial^+ + x^+D\,,
\\
\label{man-15112017-19} && K^i = -\frac{1}{2}(2x^+x^-+x^Jx^J)\partial^i +x^i D + M^{iJ} x^J + M^{i-}x^+\,,
\\
\label{man-15112017-20} && P^- = - \frac{\partial^I \partial^I}{2\partial^+} + \frac{1}{2z^2\partial^+}A\,,
\\
\label{man-15112017-21} && J^{-i}=x^-\partial^i-x^i P^- + M^{-i}\,,
\\
\label{man-15112017-22} && K^-=-\frac{1}{2}(2x^+x^- + x^I x^I) P^- + x^-D + \frac{1}{\partial^+}x^I\partial^JM^{IJ} - \frac{x^i}{2z\partial^+}[M^{zi},A] +\frac{1}{\partial^+}B\,, \qquad
\eeq
where
\be \label{man-15112017-23}
M^{-i} \equiv M^{iJ}\frac{\partial^J}{\partial^+} - \frac{1}{2z\partial^+}[M^{zi},A]\,,\qquad M^{-i}=-M^{i-}\,.
\ee
From \rf{man-15112017-14}-\rf{man-15112017-23}, we see that the differential operators are expressed in terms of space-time coordinates $x^I$, $x^\pm$, the spatial derivatives $\partial^I$, $\partial^+$, and operators $A$, $B$, $M^{IJ}$. The operators $A$, $B$, $M^{IJ}$ are independent of the space-time coordinates and the space-time derivatives. These operators act only on spin indices of the ket-vector $\phik$.
We now turn to discussion of the realization of the operators $A$, $B$, $M^{IJ}$ on the ket-vector $\phik$ \rf{man-14112017-16}.

The operators $M^{IJ} = M^{ij}, M^{zi}$ are spin operators of the $so(3)$ algebra, while the operator $M^{ij}$ is spin operator of the $so(2)$ algebra. Realization of the spin operator $M^{IJ}$ on space of ket-vector $\phik$ \rf{man-14112017-16} is well-known,
\be \label{man-15112017-24}
M^{IJ} =\alpha^I \alphab^J - \alpha^J \alphab^I\,.
\ee
In Ref.\cite{Metsaev:2003cu}, we found the following general representation for the operators $A$ and $B$:
\beq
\label{man-15112017-25} && A  = C_2 +  2 B^z + 2 M^{zi} M^{zi} + \half M^{ij}M^{ij} + \frac{15}{4}\,,
\\
\label{man-15112017-26} && B =  B^z + M^{zi} M^{zi} \,,
\eeq
where $C_2$ in \rf{man-15112017-25} stands for an eigenvalue of the 2nd-order Casimir operator of the $so(4,2)$ algebra, while $B^z$ in \rf{man-15112017-25},\rf{man-15112017-26} stands for $z$-component of a vector operator $B^I=B^i,B^z$. The operator $B^I$ acts only on spin indices of the ket-vector $\phik$ \rf{man-14112017-16} and transforms as a vector of the $so(3)$ algebra,
\be \label{man-15112017-27}
[B^I,M^{JK}] = \delta^{IJ} B^K - \delta^{IK} B^J\,.
\ee
Beside this, the operator $B^I$ should satisfy the following defining equations:
\beq
\label{man-15112017-28} && [B^I,B^J] =  (C_2 + M^2 +  4 )M^{IJ} \,,
\\
\label{man-15112017-29} && \hspace{2cm} M^2\equiv M^{IJ}M^{IJ}\,,
\eeq
where $C_2$ \rf{man-15112017-28} is an eigenvalue of the 2nd-order Casimir operator of the $so(4,2)$ algebra. It is the equations \rf{man-15112017-28} that are basic equations of light-cone gauge formulation of relativistic dynamics in $AdS_5$ space. The basic equations \rf{man-15112017-28} are the AdS counterparts of the ones in the flat space \rf{man-14112017-14}. We see that, in the flat space, the basic equations \rf{man-14112017-14} are governed by the eigenvalue of the 2nd-order Casimir operator of the Poincar\'e algebra, $C_2=m^2$, \rf{man-26112017-09},\rf{man-26112017-04}, while, in $AdS_5$ space, the basic equations \rf{man-15112017-28} are governed by the eigenvalue of the 2nd-order Casimir operator of the $so(4,2)$ algebra $C_2$ (for brief review of the Casimir operators, see Appendix A).

Finding a solution of the basic equations \rf{man-15112017-28} is the most difficult point in the framework of the light-cone gauge formulation of relativistic dynamics in $AdS_5$ space.
We find the following operator $B^I$ which satisfies Eqs.\rf{man-15112017-28}
\beq
\label{man-16112017-01} B^I & = & l S^I + g \alphab^I + A^I \gb\,,
\\
\label{man-16112017-02} && S^I \equiv \epsilon^{IJK}\alpha^J \alphab^K \,,
\\
\label{man-16112017-03} && A^I \equiv \alpha^I -  \alpha^2 \frac{1}{2N_\alpha+3} \alphab^I\,,
\\
\label{man-16112017-04} && N_\alpha \equiv \alpha^I\alphab^I\,, \qquad N_\upsilon = \upsilon \upsilonb \,,\qquad \alpha^2 =\alpha^I\alpha^I\,,
\\
\label{man-16112017-05-a1} && l \equiv \frac{\irm h_2 \kappa}{N_\upsilon (N_\upsilon +1)}\,,
\\
\label{man-16112017-05} && g \equiv g_\upsilon \upsilonb \,, \qquad  \gb =  \upsilon g_\upsilon \,, \qquad  g_\upsilon = \Bigl[\frac{((N_\upsilon+1)^2 - h_2^2)  }{(N_\upsilon+1)^3(2N_\upsilon+3)} F_\upsilon \Bigr]^{1/2}\,,
\\
\label{man-16112017-06} && F_\upsilon \equiv   \kappa^2 - (  C_2 - h_2^2 + 5)(N_\upsilon+1)^2  + (N_\upsilon+1)^4  \,,
\eeq
where $\epsilon^{IJK}$ \rf{man-16112017-02} stands for the Levi-Civita symbol of rank three with $\epsilon^{12z}=1$.
In \rf{man-16112017-05-a1},\rf{man-16112017-06}, a quantity $\kappa$ stands for a dimensionless constant parameter, while the $C_2$ is an eigenvalue of the 2nd-order Casimir operator of the $so(4,2)$ algebra. Note that the $C_2$ is also dimensionless. Thus we see that the mixed-symmetry continuous-spin field in  $AdS_5$ space is labeled by three parameters: one integer $h_2\in \No$, and two dimensionless parameters $C_2$ and $\kappa$. Helpful formulas for the operators $S^I$, $A^I$, $M^{IJ}$ may be fond in Appendix B.

The following remarks are in order.

\noindent \ibf) If $\kappa h_2 = 0$, then, using \rf{man-16112017-01},\rf{man-16112017-05-a1}, we see that the operator $B^I$ becomes real-valued. For such $\kappa$ and $h_2$, the complex-valued fields \rf{man-14112017-15} can be restricted to be real-valued and this case corresponds to totally symmetric fields.

\noindent \iibf)  If  $\kappa h_2 \ne 0$, then requiring the operator $B^I$ \rf{man-16112017-01} be hermitian, we find that $\kappa h_2$ should be real-valued, i.e., the $\kappa$ should be real-valued. To be definite let us assume that, for the ket-vector $\phik$ \rf{man-14112017-16}, the $\kappa$ is strictly positive. Thus we have the following classification:
\beq
\label{man-16112017-06-a1} && \kappa > 0 \,, \qquad \quad h_2 \in \No\,,  \hspace{2.1cm} \hbox{ for mixed-symmetry field};
\\
\label{man-16112017-06-a2}&& \kappa h_2 = 0 \,, \qquad h_2 \in \No_0\,,  \hspace{2cm} \hbox{ for totally symmetric field}.
\eeq

\noindent \iiibf) We recall that, in Sec.3, for studying the mixed-symmetry continuous-spin  field in $R^{4,1}$ space we used one integer $ h_2 \in \No $ and two {\it dimensionfull} parameters $m$ and $\kappa$, while,  in this Section, for studying the mixed-symmetry continuous-spin field  in $AdS_5$ space, we use one integer $ h_2 \in \No $ and two {\it dimensionless} parameters $C_2$ and $\kappa$.
Obviously, for the case of $AdS_5$ space, the possibility for the use of the dimensionless parameters $C_2$ and $\kappa$ is related to the radius of $AdS_5$ space. Namely, as the radius of $AdS_5$ space $R$ is a dimensionfull parameter, all dimensionfull parameters entering the game can be made dimensionless by multiplying them with a suitable powers of the $R$.

\noindent \ivbf) Using the notation $\kappa_\AdSsm$, $C_{2\AdSsm}$ and $\kappa_{\rm flat}$, $m_{\rm flat}$ for the parameters entering  the respective actions of the continuous-spin fields in AdS and flat spaces, we note that, for large $R$, these parameters are related as
\be  \label{man-20112017-03}
C_{2 \AdSsm} \bigr|_{_{R\rightarrow \infty}} \ \ \ \longrightarrow  \quad R^2  m_{\rm flat}^2  \,, \qquad \kappa_{_{\rm AdS}}\bigr|_{_{R\rightarrow \infty}} \longrightarrow\quad R \kappa_{\rm flat}\,.
\ee
Using \rf{man-20112017-03},  we note then that, for the large $R$, the operator $B^I$ \rf{man-16112017-01} in AdS space and the operator $M^I$ \rf{man-15112017-01} in flat space are related as
\be \label{man-20112017-04}
B^I \bigr|_{_{R\rightarrow \infty}} \longrightarrow  \quad  R M^I  \,.
\ee
Taking into account that the spin operator $M^{IJ}$ of the $so(3)$ algebra takes the same form in AdS space \rf{man-15112017-24} and flat space \rf{man-14112017-11}, and using \rf{man-20112017-03},\rf{man-20112017-04}, we learn that, for the large radius $R$, the basic equations \rf{man-15112017-28} in AdS space become the basic equations \rf{man-14112017-14} in flat space.

\noindent \vbf) We verify that the light-cone gauge action \rf{man-15112017-09} is invariant under the transformations of the $so(4,2)$ algebra given by
\be \label{man-20112017-01}
\delta_G |\phi\rangle = G|\phi\rangle\,,
\ee
where $G$ appearing on r.h.s \rf{man-20112017-01} stands for differential operators given in \rf{man-15112017-14}-\rf{man-15112017-22}.

\noindent \vibf) For the case of finite-component mixed-symmetry massive field in $AdS_5$ space, detailed exposition of the procedure for solving the basic equations \rf{man-15112017-28} may be found in the Appendix C in Ref.\cite{Metsaev:2004ee}. Adaptation of the procedure in the Appendix C in Ref.\cite{Metsaev:2004ee}  to the case of continuous-spin mixed-symmetry field in $AdS_5$ space is straightforward. Useful relations for various spin operators needed for the analysis of the basic equations \rf{man-15112017-28} are presented in the Appendix B in this paper.

On the one hand, the $so(4,2)$ algebra has three independent Casimir operators. On the other hand, we found that the Lagrangian of the mixed-symmetry continuous-spin field in  $AdS_5$ space depends on three parameters: one integer $h_2\in \No$, one dimensionless real-valued parameter $\kappa$ and eigenvalue of the 2nd-order Casimir operator $C_2$. Our aim is to express eigenvalues of 3rd-order and 4th-order Casimir operators of the $so(4,2)$ algebra in terms of the $h_2$, $\kappa$, and $C_2$. To this end we now present our new results for the light-cone gauge representation for 3rd-order and 4th-order Casimir operators of the $so(4,2)$ algebra which has not been discussed in Refs.\cite{Metsaev:2003cu,Metsaev:2004ee}.

\noindent {\bf Casimir operators of the $so(4,2)$ algebra}. We find that 3rd-order and 4th-order Casimir operators of the $so(4,2)$ algebra, denoted as $C_{\epsilon\, 3}$ and $C_4$ respectively, can be expressed in terms of eigenvalues of the 2nd-order Casimir operator $C_2$ and the spin operators $B^I$, $M^{IJ}$ as follows
\beq
\label{man-26112017-15} && C_{\epsilon\, 3} = \frac{\irm }{2} \epsilon^{ IJK} B^I M^{JK},
\\
\label{man-26112017-16} && C_4 = B^IB^I  -  \half ( C_2 + 2 ) M^2 - \frac{1}{4} (M^2 )^2 \,, \qquad M^2\equiv M^{IJ}M^{IJ}\,.
\eeq
The statement that operators $C_{\epsilon\, 3}$, $C_4$ \rf{man-26112017-15},\rf{man-26112017-16} are indeed Casimir operators of the $so(4,2)$ algebra can directly be verified by using \rf{man-15112017-27},\rf{man-15112017-28}. See also helpful relations in Appendix B. For the reader convenience, in Appendix A, we briefly review a manifestly 6-dimensional covariant representation for the Casimir operators of the $so(4,2)$ algebra.

Plugging the operators $M^{IJ}$, $B^I$ \rf{man-15112017-24}, \rf{man-16112017-01} into \rf{man-26112017-15},\rf{man-26112017-16}, we find that the operators $C_{\epsilon\, 3}$, $C_4$  are diagonalized,
\beq
\label{man-26112017-17} && C_{\epsilon\, 3}  = \kappa h_2\,,
\\
\label{man-26112017-18} && C_4 =  \kappa^2 + (C_2 - h_2^2 + 4)(h_2^2-1)\,.
\eeq
From \rf{man-26112017-17},\rf{man-26112017-18}, we see how the eigenvalues of the Casimir operators $C_{\epsilon\, 3}$, $C_4$ are expressed in terms of $\kappa$, $h_2$, and $C_2$. Note that, eigenvalues of the $C_2$ for the ket-vector $\phik$ and the bra-vector $\phibr$, $\phibr\equiv \phik^\dagger$, are equal. The same holds true for eigenvalues of the $C_4$. Contrary this, eigenvalue of the $C_{\epsilon,3}$ \rf{man-26112017-15} for the ket-vector $\phik$ is equal to $\kappa h_2$ \rf{man-26112017-17}, while, eigenvalue of the $C_{\epsilon,3}$ \rf{man-26112017-15} for the bra-vector $\phibr$ is equal to $-\kappa h_2$.  By definition, the ket-vector $\phik$ and the bra-vector $\phibr$ have one and the same label $h_2$. This implies that ket-vector $\phik$ \rf{man-14112017-16} is related to the representation of the $so(4,2)$ algebra labeled by $C_2$, $\kappa$, $h_2$, while the bra-vector $\phibr$ is related to the representation of the $so(4,2)$ algebra labeled by $C_2$, $-\kappa$, $h_2$.

\noindent {\bf Totally symmetric continuous-spin field}. From \rf{man-16112017-06-a2}, we learn that totally symmetric continuous-spin field is realized by considering the following two cases:
\beq
\label{man-10022018-01}  && h_2  = 0 \,, \hspace{1cm} \kappa -\hbox{arbitrary} \,,
\\
\label{man-10022018-02} && h_2  \ne 0 \,, \hspace{1cm} \kappa = 0 \,.
\eeq

\noindent {\bf Case $h_2=0$, $\kappa$-arbitrary}. Setting $h_2=0$ in \rf{man-14112017-16}, we get ket-vector $\phik$ entering Lagrangian for totally symmetric field \rf{man-15112017-10}. Also, setting $h_2=0$ in \rf{man-16112017-01}-\rf{man-16112017-06}, we see that the operator $B^I$ is simplified as
\beq
\label{man-10022018-03} && B^I =  g \alphab^I + A^I \gb\,,
\\
\label{man-10022018-04}  &&   g = g_\upsilon \upsilonb \,, \qquad \gb =  \upsilon g_\upsilon \,, \qquad  g_\upsilon = \Bigl( \frac{ F_\upsilon }{(N_\upsilon+1) (2N_\upsilon+3)} \Bigr)^{1/2}\,,
\\
\label{man-10022018-05} &&F_\upsilon \equiv   \kappa^2 - (  C_2  + 5)(N_\upsilon+1)^2  + (N_\upsilon+1)^4  \,,
\eeq
where the operators $A^I$, $N_\upsilon$ are defined in \rf{man-16112017-02},\rf{man-16112017-03}.

The following remarks are in order:

\noindent \ibf) As the $lS^I$-term \rf{man-16112017-01} does not appear in \rf{man-10022018-03},  the hermitian operator $B^I$ \rf{man-10022018-03} turns out to be real-valued. Therefore the complex-valued fields \rf{man-14112017-15} can be restricted to be real-valued.

\noindent \iibf) We recall that, for the mixed-symmetry field, the hermicity of the operator $B^I$ \rf{man-16112017-01} implies, in view of the $lS^I$-term \rf{man-16112017-01}, that the $\kappa$ should be real-valued. For the totally symmetric field, the $lS^I$-term \rf{man-16112017-01} does not appear in \rf{man-10022018-03}. Therefore, for the totally symmetric field, the hermicity of the operator $B^I$ \rf{man-10022018-03} does not imply that only real-valued $\kappa$ are admitted. Namely, from \rf{man-10022018-03}-\rf{man-10022018-05}, we see that the hermicity of the operator $B^I$ implies that $\kappa^2$ should be real-valued. In other words, for the totally symmetric field, the $\kappa$ can be real-valued or purely imaginary. For more discussion on this, see Sec.4.

\noindent {\bf Case $h_2 \ne 0$, $\kappa=0$}. In Sec.4, we will demonstrate that this case is realized as some infinite-component field entering reducible continuous-spin field that has  $h_2=0$, $\kappa\ne 0 $.

%%%%%%%%%%%%%%%%%%%%%%%%%%%%%%%%%%%%%%%%%%%%%%%%%%%%%%%%%%%%%%%%%%%%%%%%%%%%%%%%%%%%%%%%%%%
%%%%%%%%%%%%%%%%%%%%%%%%%%%%%%%%%%%%%%%%%%%%%%%%%%%%%%%%%%%%%%%%%%%%%%%%%%%%%%%%%%%%%%%%%%%
\newsection{\large  (Ir)reducible classically unitary mixed-symmetry continuous-spin \\ field in $AdS_5$ space}\label{sec-04}
%%%%%%%%%%%%%%%%%%%%%%%%%%%%%%%%%%%%%%%%%%%%%%%%%%%%%%%%%%%%%%%%%%%%%%%%%%%%%%%%%%%%%%%%%%%
%%%%%%%%%%%%%%%%%%%%%%%%%%%%%%%%%%%%%%%%%%%%%%%%%%%%%%%%%%%%%%%%%%%%%%%%%%%%%%%%%%%%%%%%%%%

Lagrangian of the mixed-symmetry continuous-spin field in $AdS_5$ \rf{man-15112017-10} depends on the three parameters $C_2$, $\kappa$ and $h_2$. We now discuss restrictions imposed on these parameters for irreducible and reducible classically unitary dynamical systems.  We start with our definition of classically unitary reducible and irreducible systems.

Light-cone gauge Lagrangian \rf{man-15112017-10} is constructed out of complex-valued fields. In order for the light-cone gauge action be real-valued the parameter $\kappa$ \rf{man-16112017-05-a1} entering operator $B^I$ \rf{man-16112017-01} should be real-valued, while the $F_\upsilon$ defined in \rf{man-16112017-06} should be positive for all eigenvalues $N_\upsilon=h_2,h_2+1,\ldots,\infty$. Introducing the notation
\beq
\label{man-26112017-19} F_\upsilon(n) & =  &  \kappa^2 - \mu (n+1)^2  + (n+1)^4\,, \qquad  F_\upsilon(n) \equiv F_\upsilon|_{N_\upsilon=n}\,,
\\
\label{man-26112017-20} && \mu \equiv   C_2 - h_2^2 + 5\,,
\eeq
we note that, depending on behaviour of the $F_\upsilon(n)$, we use the following terminology
\beq
\label{man-26112017-21} && F_\upsilon(n)\geq 0 \ \hbox{ for all } \ n=h_2,h_2+1,\ldots, \infty  \hspace{1.8cm} \hbox{ classically unitary system}; \qquad
\\
\label{man-26112017-22} && F_\upsilon(n) \ne 0 \ \hbox{ for all } \ n =  h_2,h_2+1,\ldots, \infty,  \hspace{1.7cm} \hbox{ irreducible system};
\\
\label{man-26112017-23} && F_\upsilon(n_r) = 0 \ \hbox{ for some } \ n_r \in h_2,h_2+1,\ldots, \infty,  \hspace{1cm} \hbox{ reducible system}.
\eeq
This is to say that, if $F_\upsilon(n)$ \rf{man-26112017-19} is positive for all $n$  \rf{man-26112017-21}, then we will refer to fields \rf{man-14112017-16} as classically unitary system, while if $F_\upsilon(n)$ \rf{man-26112017-19} has no roots \rf{man-26112017-22},  then we will refer to fields \rf{man-14112017-16} as irreducible system. For the case \rf{man-26112017-22}, our Lagrangian \rf{man-15112017-10} describes infinite chain of coupling fields \rf{man-14112017-16}. Relation \rf{man-26112017-23} tells us that, if $F_\upsilon(n)$ \rf{man-26112017-19} has roots, then we will refer to fields \rf{man-14112017-16} as reducible system. For the case of the reducible system, Lagrangian \rf{man-15112017-10} is factorized and describe finite and infinite decoupled chains of fields.

Taking into account the definitions presented in \rf{man-26112017-21}-\rf{man-26112017-23}, we now define (ir)reducible classically unitary systems in the following way:
{\small
\beq
\label{man-26112017-24}  && \hspace{-1.2cm} F_\upsilon(n) > 0 \ \ \ \hbox{ for all } \ n =h_2,h_2+1,\ldots, \infty,  \hspace{2.3cm} \hbox{ irreducible classically unitary system};\qquad
\\
&& \hspace{-1.2cm}  F_\upsilon(n_r) = 0  \ \ \hbox{ for some } n_r \in  h_2,h_2+1,\ldots, \infty\,,
\nonumber\\
\label{man-26112017-25} && \hspace{-1.2cm} F_\upsilon(n) > 0  \ \ \ \hbox{ for all } n =h_2,h_2+1,\ldots, \infty \hbox{ and } n \ne n_r \hspace{0.6cm} \hbox{ reducible classically unitary system}.
\eeq
}
For the mixed-symmetry field, the $\kappa$ associated with the ket-vector $\phik$ is strictly positive, $\kappa>0$ \rf{man-16112017-06-a1}. Keeping this in mind, we now summarize our study of Eqs.\rf{man-26112017-24},\rf{man-26112017-25} as the following three Statements.

\noindent {\bf Statement 1}. We classify solutions of Eqs.\rf{man-26112017-24} as type $I$, $IIA$, $IIB$, and $III$ solutions. These solutions are as follows.
\beq
\label{man-27112017-16} && \hspace{-2.5cm} \hbox{\it Type I solutions:}
\nonumber\\
&&\mu < 2\kappa; \quad \kappa > 0\,,
\\
&& \hspace{-2.5cm} \hbox{\it Type IIA solutions:}
\nonumber\\
\label{man-27112017-17} && \mu > 2\kappa\,,\qquad \kappa > 0
\\
\label{man-27112017-18} && \mu = p^2 + q^2 \,, \qquad \kappa = pq\,,
\\
\label{man-27112017-19} && 0 <  p < h_2 + 1\,, \qquad 0 < q < h_2+1; \qquad p\ne q;
\\
&& \hspace{-2.5cm} \hbox{\it Type IIB solutions:}
\nonumber\\
\label{man-27112017-20} && \mu > 2\kappa\,, \qquad \kappa > 0\,,
\\
\label{man-27112017-21} && \mu = p_k^2 + q_k^2 \,, \qquad \kappa = p_k q_k\,,
\\
\label{man-27112017-22} && p_k = h_2 + 1+ k + \epsilon_p\,,  \qquad q_k = h_2 + 1 + k + \epsilon_q\,, \qquad k\in \No_0\,.\qquad
\\
\label{man-27112017-23} && 0 < \epsilon_p < 1\,, \qquad 0< \epsilon_q<1\,, \qquad \epsilon _p \ne \epsilon_q\,.
\\
&& \hspace{-2.5cm} \hbox{\it Type III solutions:}
\nonumber\\
\label{man-27112017-24} && \mu = 2\kappa\,, \qquad \kappa > 0\,,
\\
\label{man-27112017-25} && \mu = 2 p_k^2 \,, \qquad \kappa = p_k^2\,,
\\
\label{man-27112017-26} && p_k =   k + \epsilon_p\,,  \qquad   0 < \epsilon_p < 1\,, \qquad k \in \No_0\,.\qquad
\eeq
We now make comments on the Statement 1.

\noindent \ibf) For the type I solutions, using \rf{man-26112017-20},\rf{man-27112017-16}, we get the following restriction for the eigenvalue of the 2nd-order Casimir operator $C_2$:
\be
C_2 < 2\kappa + h_2^2 - 5\,.
\ee

\noindent \iibf) For the type II and III solutions, we can get various interesting representations for the eigenvalues of the Casimir operators. Namely, using \rf{man-26112017-20} and \rf{man-27112017-18}, \rf{man-27112017-21}, \rf{man-27112017-25}, we see that $C_2$ and $C_{\epsilon\, 3}$, $C_4$ \rf{man-26112017-17},\rf{man-26112017-18} can be represented as
\beq
\label{man-28112017-15} && C_2 = p^2 + q^2 + h_2^2 - 5\,,
\\
\label{man-28112017-16} && C_{\epsilon\, 3}  = pq  h_2\,,
\\
\label{man-28112017-17} && C_4  = (p^2-1)(q^2-1) +  h_2^2 (p^2 + q^2 -1)\,,
\eeq
where, for the type $IIB$ solutions, values of the $p$, $q$ are given in \rf{man-27112017-22}.
Note that, for the type III solutions, we should set $p=q$ in \rf{man-28112017-15}-\rf{man-28112017-17}, where values of the $p$ are given in \rf{man-27112017-26}.

\noindent \iiibf) Other interesting representation for the eigenvalues of the Casimir operators is obtained by using, in place of the $p$, $q$, and $h_2$, new parameters $\EE_0$, $\HH_1$, $\HH_2$ defined by the relations
\beq
\label{man-28112017-18} && p = \EE_0-2\,,
\\
\label{man-28112017-19} && q = \HH_1 + 1\,,
\\
\label{man-28112017-20} && h_2 = \HH_2\,.
\eeq
Plugging \rf{man-28112017-18}-\rf{man-28112017-20} into \rf{man-28112017-15}-\rf{man-28112017-17},
we find the following expressions:
\beq
&& C_2 = \EE_0(\EE_0-4) + \HH_1(\HH_1+2) + \HH_2^2\,,
\nonumber\\
\label{man-28112017-22} && C_{\epsilon\, 3} = (\EE_0-2)(\HH_1+1)\HH_2\,,
\\
&& C_4 = (\EE_0-1)(\EE_0-3) \HH_1(\HH_1+2) + \HH_2^2 \Bigl(\EE_0(\EE_0-4) + \HH_1 (\HH_1+2) + 4 \Bigr)\,.\qquad
\nonumber
\eeq
We now explain our motivation for introducing the parameters $\EE_0$, $\HH_1$, $\HH_2$.
To this end we use the notation $D(E_0,h_1,h_2)$ for a positive-energy lowest weight representation of the $so(4,2)$ algebra, where $E_0$ is a lowest eigenvalue of the energy operator, while $h_1$, $h_2$ label the highest weight of the $so(4)$ algebra representation. Eigenvalues of the Casimir operators for the $D(E_0,h_1,h_2)$ are given in Appendix A in \rf{man-28112017-12}.
We see then that relations in \rf{man-28112017-22} are similar to the ones in \rf{man-28112017-12}. It is the similarity of relations in \rf{man-28112017-22} and \rf{man-28112017-12} that motivates us to introduce the parameters $\EE_0$, $\HH_1$, $\HH_2$ in \rf{man-28112017-18}-\rf{man-28112017-20}.

\noindent \ivbf) For the type II and III solutions, the classical unitarity restrictions imposed on the parameters $p$, $q$, $h_2$ can straightforwardly  be expressed in terms of the $\EE_0$, $\HH_1$, $\HH_2$ defined in \rf{man-28112017-18}-\rf{man-28112017-20}. Namely, for the type $IIB$ solutions, we get the restrictions
\beq
&&  \EE_0 >  \HH_1 + 2\,, \qquad \HH_1  > \HH_2\,,
\nonumber\\[-10pt]
\label{man-28112017-27}  && \hspace{10cm} \hbox{ for IIB solutions}\qquad
\\[-10pt]
&& \HH_2 + 3 <  \EE_0 < \HH_1 + 4\,, \qquad \EE_0 \ne \HH_1 + 3\,,
\nonumber
\eeq
while, for the type $IIA$ and $III$ solutions, we get the restrictions
\beq
\label{man-07022018-01}  &&  \hspace{-1cm}  2 < \EE_0 <  \HH_2 + 3\,, \qquad -1 < \HH_1  < \HH_2\,,  \qquad \EE_0 \ne \HH_1 + 3\,, \hspace{0.6cm} \hbox{ for IIA solutions}; \qquad
\\
\label{man-07022018-02} && \hspace{-1cm}  \EE_0 =  \HH_1 + 3\,, \hspace{1.6cm} -1 < \HH_1\,,  \hspace{5cm} \hbox{ for III solutions}.
\eeq
Note that restrictions \rf{man-07022018-01} are easily obtained by using the classical unitarity restrictions \rf{man-27112017-19} and relations \rf{man-28112017-18}-\rf{man-28112017-20}. To get \rf{man-07022018-02}, we use the $p>0$ \rf{man-27112017-26}, and make the identification $p=q$ in \rf{man-28112017-18},\rf{man-28112017-19}.
To get \rf{man-28112017-27} we note that, for the type $IIB$ solutions, the classical unitarity restrictions given in \rf{man-27112017-22},\rf{man-27112017-23} amount to the restrictions
\be \label{man-28112017-26}
-1 < p - q < 1\,, \qquad p > h_2 + 1\,, \qquad q > h_2 + 1\,,  \qquad p \ne q\,,\hspace{0.5cm} \hbox{ for IIB solutions} \,.
\ee
Using \rf{man-28112017-18}-\rf{man-28112017-20}, we see that restrictions in \rf{man-28112017-26} amount to restrictions in \rf{man-28112017-27}.

\noindent \vbf) As we noted above, the parameters $E_0$, $h_1$, $h_2$ label the positive-energy lowest weight representation of the $so(4,2)$ algebra denoted by $D(E_0,h_1,h_2)$.  We now compare unitarity restrictions imposed on the $E_0$, $h_1$, $h_2$ with the classical unitarity restrictions imposed on the $\EE_0$, $\HH_1$, $\HH_2$ in \rf{man-28112017-27}. To this end we recall that, if $h_1>h_2$, then the $D(E_0,h_1,h_2)$ is realized as unitary representation of the $so(4,2)$ algebra provided the $E_0$, $h_1$, $h_2$ satisfy the following restrictions:
\be \label{man-28112017-24}
E_0 \geq h_1 +2 \,, \qquad \hbox{ for } \ h_1 > h_2 \,.
\ee

The restrictions in the 1st line in \rf{man-28112017-27} are remarkably similar to the ones in \rf{man-28112017-24}. Note however that the label $h_1$ in \rf{man-28112017-24} is integer, while our label $\HH_1$ is not integer. Also note that, contrary to \rf{man-28112017-24}, for the type IIB solutions, we have the additional restrictions given in the 2nd line in \rf{man-28112017-27}.

\noindent {\bf Statement 2}. Solution to Eqs.\rf{man-26112017-25} with one root of $F_\upsilon$ denoted by $h_1$ is given by
\beq
\label{man-26112017-26} && \kappa^2 = (h_1+1)^2 \left(C_2 - h_1(h_1+2) - h_2^2 +4\right),
\\
\label{man-26112017-27} && 2h_1(h_1+1) + h_2^2 - 4 < C_2 < 2h_1(h_1+3) + h_2^2\,,
\\
\label{man-26112017-27-a1} && F_\upsilon =  \Bigl( (h_1+1)^2 -(N_\upsilon+1)^2 \Bigr) \Bigl(  C_2 - h_1(h_1+2) -  h_2^2  + 4 - (N_\upsilon+1)^2 \Bigr)\,.\qquad
\eeq
We now make comments on the Statement 2.

\noindent \ibf) Lagrangian \rf{man-15112017-10}, with the $C_2$ and $F_\upsilon$ as in \rf{man-26112017-27}, \rf{man-26112017-27-a1}, describes a reducible classically unitary system. This is to say that decomposing  ket-vector $\phik$ \rf{man-14112017-16} as
\beq
\label{man-27112017-01} \phik & = &  |\phi^{h_2,h_1}\rangle  + |\phi^{h_1+1,\infty}\rangle \,,
\\
\label{man-27112017-02} && |\phi^{M,N}\rangle \equiv  \sum_{n=M}^N \frac{\upsilon^n}{n!\sqrt{n!}} \alpha^{I_1} \ldots \alpha^{I_n} \phi^{I_1\ldots I_n}(x,z)|0\rangle\,,
\eeq
we verify that Lagrangian \rf{man-15112017-10} is factorized as
\beq
\label{man-27112017-03} \LL & = &  \LL^{h_2,h_1} + \LL^{h_1+1,\infty}\,,
\\
\label{man-27112017-04} && \LL^{M_,N}  \equiv   \langle \phi^{M,N} |\bigl(\Box - \frac{1}{z^2}A \bigr) |\phi^{M,N} \rangle\,.
\eeq
Namely, $\LL^{h_2,h_1}$ and $\LL^{h_1+1,\infty}$ \rf{man-27112017-03} are invariant under the $so(4,2)$ algebra transformations \rf{man-20112017-01}.

\noindent \iibf) The $|\phi^{h_2,h_1}\rangle$ \rf{man-27112017-01} describes massive finite-component field associated with positive-energy lowest weight representation of the $so(4,2)$ algebra which we denote as $D(E_0,h_1,h_2)$. Eigenvalue of the 2nd-order Casimir operator $C_2$, the labels $E_0$, $h_1$, $h_2$ and a mass parameter $m^2$ for $|\phi^{h_2,h_1}\rangle$ are related as
\beq
\label{man-27112017-27} &&  C_2 = E_0(E_0-4) + h_1(h_1+2) + h_2^2\,,
\\
\label{man-04022017-03}  && m^2 \equiv (E_0-2)^2 - h_1^2\,.
\eeq
Using \rf{man-27112017-27}, we can represent relations \rf{man-26112017-26},\rf{man-26112017-27} as
\beq
\label{man-26112017-32} && \kappa = (E_0-2)(h_1+1)\,,
\\
\label{man-26112017-33} && h_1 + 2 < E_0  < h_1 + 4\,.
\eeq
In turn, using the mass parameter \rf{man-04022017-03}, we can represent relation \rf{man-26112017-33} as
\be
0 < m^2 < 4 (h_1+1)\,.
\ee

\noindent \iiibf) For a particular $C_2$ in \rf{man-26112017-27-a1}, the root $h_1$ becomes doubly-degenerate. Namely, appearance of the doubly-degenerate root $h_1$ in \rf{man-26112017-27-a1} implies the following relations for $\kappa$, $C_2$, and $F_\upsilon$:
\beq
\label{man-05022017-01} && \kappa = (h_1+1)^2\,,
\\
\label{man-05022017-02} && C_2 = 2h_1(h_1+2) + h_2^2 -3\,,
\\
\label{man-05022017-02-a1} && F_\upsilon =  \Bigl( (h_1+1)^2 -(N_\upsilon+1)^2 \Bigr)^2\,.
\eeq
Lagrangian \rf{man-15112017-10}, with $F_\upsilon$ as in \rf{man-05022017-02-a1}, describes a reducible classically unitary system. Namely, for $F_\upsilon$ given in \rf{man-05022017-02-a1}, the ket-vector $\phik$ and the Lagrangian are decomposed as in \rf{man-27112017-01} and \rf{man-27112017-03} respectively. The $|\phi^{h_2,h_1}\rangle$ \rf{man-27112017-01} describes classically unitary finite-component massive field, while the $|\phi^{h_1+1,\infty}\rangle$ \rf{man-27112017-01} describes classically unitary infinite-component spin field. Lowest eigenvalue of the energy operator and mass parameter $m^2$ of the $|\phi^{h_2,h_1}\rangle $ are given by
\be \label{man-05022017-06}
E_0 = h_1 + 3 \,, \qquad m^2 = 2h_1 + 1\,, \hspace{3cm} \hbox{ for massive} \ \ |\phi^{h_2,h_1}\rangle\,.\qquad
\ee
Note that $E_0$ \rf{man-05022017-06} is obtained by plugging \rf{man-05022017-02} into \rf{man-27112017-27}. In turn, plugging $E_0$ \rf{man-05022017-06} into \rf{man-04022017-03}, leads to $m^2$ given in \rf{man-05022017-06}.

\noindent {\bf Statement 3}. Solution to Eqs.\rf{man-26112017-25} with two roots of $F_\upsilon$ is given by
\beq
\label{man-04022017-01} && \kappa  = (h_1+1)(h_1+2)\,,
\\
\label{man-04022017-02} && C_2 = 2h_1(h_1+3) + h_2^2\,,
\\
\label{man-04022017-02-a1} && F_\upsilon =  \Bigl( (h_1+1)^2 -(N_\upsilon+1)^2 \Bigr)\Bigl( (h_1+2)^2 -  (N_\upsilon+1)^2 \Bigr)\,.
\eeq
We now make comments on the Statement 3.

\noindent \ibf) From \rf{man-04022017-02-a1}, we see that the $F_\upsilon$ has two roots. Namely, the $F_\upsilon$ is equal to zero for $N_\upsilon=h_1$ and $N_\upsilon = h_1+1$.

\noindent \iibf) Lagrangian \rf{man-15112017-10}, with $F_\upsilon$ as in \rf{man-04022017-02-a1}, describes a reducible classically unitary field. Namely, using  $F_\upsilon$ given in \rf{man-04022017-02-a1} and the notation given in \rf{man-27112017-02},\rf{man-27112017-04}, we can decompose $\phik$ \rf{man-14112017-16} and Lagrangian \rf{man-15112017-10} as
\beq
\label{man-05022017-07} && \phik = |\phi^{h_2,h_1}\rangle  + |\phi^{h_1+1,h_1+1}\rangle + |\phi^{h_1+2,\infty}\rangle \,,
\\
\label{man-05022017-08} && \LL = \LL^{h_2,h_1}  + \LL^{h_1+1,h_1+1} + \LL^{h_1+2,\infty} \,.
\eeq

\noindent \iiibf) In \rf{man-05022017-07}, the $|\phi^{h_2,h_1}\rangle$ describes classically unitary finite-component massive field associated with the representation $D(E_0,h_1,h_2)$, the $|\phi^{h_1+1,h_1+1}\rangle$  describes classically unitary finite-com\-po\-nent massless field associated with the representation $D(E_0,h_1+1,h_2)$, while the $|\phi^{h_1+2,\infty}\rangle$ describes classically unitary infinite-component spin field. An lowest eigenvalues of the energy operator and mass parameters $m^2$ of the fields $|\phi^{h_2,h_1}\rangle$ and $|\phi^{h_1+1,h_1+1}\rangle $ are given by
\beq
\label{man-05022017-09} && E_0 = h_1 + 4\,, \qquad m^2 = 4(h_1+1) \,, \hspace{1.6cm} \hbox{ for massive } \ \ |\phi^{h_2,h_1}\rangle\,,
\\
\label{man-05022017-10} && E_0 = h_1 + 3\,, \qquad m^2 = 0 \,, \hspace{3cm} \hbox{ for massless } \ \ |\phi^{h_1+1,h_1+1}\rangle\,.\qquad
\eeq

The three Statements above-presented can easily be proved by noticing that $F_\upsilon(n)$ \rf{man-26112017-19} has at most two roots. This is to say that we are going to analyse the following three cases: 1) $F_\upsilon(n)$ has no roots; 2) $F_\upsilon(n)$ has one root; 3) $F_\upsilon(n)$ has two roots. Let us analyse these three cases in turn.

\noindent \ibf) {\bf Solutions without roots of $F_\upsilon(n)$}. Such solutions lead to the type $I$, $II$, and $III$ solutions above-described in the Statement 1. We consider them separately.

First, using \rf{man-26112017-19}, we note that, if $\mu < 2\kappa$, $\kappa>0$, then $F_\upsilon(n)>0$ for all $n$ \rf{man-26112017-24}. This gives type I solutions in \rf{man-27112017-16}.

Second, we consider the case $\mu> 2\kappa$, $\kappa>0$. Such $\mu$ and $\kappa$ can be presented in terms of two positive non-equal numbers $p$, $q$ as
\be \label{man-26112017-28}
\mu = p^2 + q^2 \,, \quad \kappa = p q\,, \qquad p \ne q \,, \qquad p> 0 \,, \qquad q >0\,,
\ee
Plugging $\mu$ and $\kappa$ \rf{man-26112017-28} into \rf{man-26112017-19}, we represent $F_\upsilon(n)$ as %
\be \label{man-26112017-29}
F_\upsilon(n) = \Bigl( (n+1)^2 - p^2 \Bigr) \Bigl( (n+1)^2 - q^2 \Bigr)\,.
\ee
Using \rf{man-26112017-29}, it is easy to see that Eqs.\rf{man-26112017-24} lead to the type IIA solutions \rf{man-27112017-17}-\rf{man-27112017-19} and the type IIB solutions \rf{man-27112017-20}-\rf{man-27112017-23}.

Third, we consider the case $\mu = 2\kappa$, $\kappa>0$. Then we can represent $\mu$, $\kappa$, and $F_\upsilon(n)$ as
\be \label{man-26112017-30}
\mu = 2p^2 \,, \qquad \kappa = p^2\,, \qquad p> 0 \,, \qquad F_\upsilon(n) = ((n+1)^2 - p^2)^2\,.
\ee
Using $F_\upsilon(n)$ \rf{man-26112017-30}, it is easy to see that Eqs.\rf{man-26112017-24} lead to the type III solutions in \rf{man-27112017-24}-\rf{man-27112017-26}.

Finishing discussion of the Statement 1 we explain our two motivations for introducing the parameters $p$ and $q$ \rf{man-26112017-28}. First, the use of such parameters allows us to obtain factorized representation for $F_\upsilon$ given in \rf{man-26112017-29} which turns out to be very convenient for the analysis of the requirement of the classical unitarity. Second, in terms of the $p$ and $q$, the eigenvalues of the Casimir operators $C_2$, $C_{\epsilon,3}$, $C_4$ take simple and convenient form given in \rf{man-28112017-15}-\rf{man-28112017-17}.

\noindent \iibf) {\bf Solution with one root of $F_\upsilon(n)$}.  Solutions of Eqs.\rf{man-26112017-25} with one root of $F_\upsilon(n)$ are described in the Statement 2.

First, we outline the derivation of relations in \rf{man-26112017-26}-\rf{man-26112017-27-a1}. Denoting one root of $F_\upsilon$ as $h_1$, we see that Eqs.\rf{man-26112017-25} can be represented as
\be \label{man-27112017-05}
F_\upsilon(h_1) = 0\,, \qquad  F_\upsilon(n) > 0 \quad \hbox{ for } \quad n =h_2, h_2+1,\ldots, h_1-1, h_1+1,h_1+2,\ldots, \infty\,.
\ee
Using \rf{man-26112017-19}, we note then that the equation $F_\upsilon(h_1) = 0$ leads to the following value of $\kappa^2$:
\be \label{man-27112017-06}
\kappa^2 = (h_1+1)^2 \left(C_2 - h_1(h_1+2) - h_2^2 +4\right)\,.
\ee
Plugging $\kappa^2$ \rf{man-27112017-06}  into \rf{man-16112017-06}, we get $F_\upsilon$ given in \rf{man-26112017-27-a1}, while plugging $\kappa^2$ \rf{man-27112017-06}  into \rf{man-26112017-19}, we cast the $F_\upsilon(n)$ into the following factorized form:
\be \label{man-27112017-07}
F_\upsilon(n) =  \Bigl( (h_1+1)^2 -(n+1)^2 \Bigr)\Bigl(  C_2 - h_1(h_1+2) -  h_2^2   - (n+1)^2 + 4 \Bigr)\,.
\ee
Now, using $F_\upsilon(n)$ \rf{man-27112017-07}   and considering inequalities $F_\upsilon(n) > 0$ in \rf{man-26112017-25}, we find the restrictions on the $C_2$,
\be \label{man-27112017-08}
2h_1(h_1+1) + h_2^2 - 4 < C_2 < 2h_1(h_1+3) + h_2^2\,.
\ee
For the reader convenience, we note that the left inequality in \rf{man-27112017-08}  is obtained by requiring $F_\upsilon(n)>0$ for $n=h_2,h_2+1,\ldots,h_1-1$, while the right inequality in \rf{man-27112017-08} is obtained by requiring $F_\upsilon(n)>0$ for $n=h_1+1,h_1+2,\ldots,\infty$.
Note that using the left inequality \rf{man-27112017-08} in \rf{man-27112017-06}, we find $\kappa^2>0$, as it should be for the real-valued $\kappa$.

Second, we outline the derivation of relations in \rf{man-05022017-01}-\rf{man-05022017-02-a1}. From \rf{man-27112017-07}, we see that, if $C_2$ takes the value
\be \label{man-27112017-08-a1}
C_2 = 2h_1(h_1+2) + h_2^2 -3\,,
\ee
then the root $h_1$ in \rf{man-27112017-07} becomes double-degenerate. Plugging $C_2$ \rf{man-27112017-08-a1} into \rf{man-27112017-06}, we get $\kappa$ given in \rf{man-05022017-01}, while plugging $C_2$ \rf{man-27112017-08-a1} into \rf{man-26112017-27-a1}, we get $F_\upsilon$ given in \rf{man-05022017-02-a1}.

\noindent \iiibf) {\bf Solution with two roots of $F_\upsilon(n)$}.  Solutions of Eqs.\rf{man-26112017-25} with two roots of $F_\upsilon(n)$ are described in the Statement 3.  Denoting two roots of $F_\upsilon$ as $h_1$ and $H_1$,
\be  \label{man-27112017-09}
F_\upsilon(h_1) = 0 \,, \qquad F_\upsilon(H_1) = 0\,, \qquad h_1 < H_1\,,
\ee
we see that Eqs.\rf{man-27112017-09}  lead to the following relations:
\beq
\label{man-27112017-10}   && \kappa =  ( h_1 + 1)  (H_1+1)\,,
\\
\label{man-27112017-11}  && C_2  = H_1(H_1+2) + h_1(h_1+2) + h_2^2 - 3\,.
\eeq
Plugging $\kappa$, $C_2$ \rf{man-27112017-10},\rf{man-27112017-11} into \rf{man-26112017-19}, we find the following factorized representation for $F_\upsilon(n)$:
\be \label{man-27112017-12}
F_\upsilon(n) =  \Bigl( (h_1+1)^2 -(n+1)^2 \Bigr)\Bigl( (H_1+1)^2 -  (n+1)^2 \Bigr)\,.
\ee
Lagrangian \rf{man-15112017-10}, with $\kappa$ and $C_2$ as in \rf{man-27112017-10},\rf{man-27112017-11},  describes reducible mixed-symmetry conti\-nuous-spin field.  Namely, using $\kappa$ and $C_2$ given in \rf{man-27112017-10}, \rf{man-27112017-11} and the notation given in \rf{man-27112017-02}, \rf{man-27112017-04}, we can decompose ket-vector $\phik$ \rf{man-14112017-16} and Lagrangian \rf{man-15112017-10} as
\beq
\label{man-27112017-14}  && \phik = |\phi^{h_2,h_1}\rangle  + |\phi^{h_1+1,H_1}\rangle + |\phi^{H_1+1,\infty}\rangle \,,
\\
\label{man-27112017-15} && \LL = \LL^{h_2,h_1} + \LL^{h_1+1,H_1} + \LL^{H_1+1,\infty}\,.
\eeq
Let us consider the cases $h_1+1=H_1$, and $h_1 + 1 < H_1$ separately.

\noindent {\bf Case $h_1+1=H_1$}. It is this case that respects the classical unitarity. This case is described in the Statement 3. Namely, setting $h_1+1=H_1$ in \rf{man-27112017-12}, we see that the $|\phi^{h_2,h_1}\rangle$ describes classically unitary massive finite-component field, the $|\phi^{h_1+1,h_1+1}\rangle$ describes classically unitary massless finite-component field, while the  $|\phi^{h_1+2,\infty}\rangle$ describes classically unitary infinite-component spin field.
Setting $h_1+1=H_1$ in \rf{man-27112017-10},\rf{man-27112017-11}, we find the $\kappa$ and $C_2$ given in \rf{man-04022017-01},\rf{man-04022017-02}. In turn, using $C_2$ \rf{man-04022017-02}, we find then the $E_0$ and $m^2$ of the fields $|\phi^{h_2,h_1}\rangle$ and $|\phi^{h_1+1,h_1+1}\rangle$ given in \rf{man-05022017-09},\rf{man-05022017-10}. Plugging $\kappa$ \rf{man-04022017-01} and $C_2$ \rf{man-04022017-02} into \rf{man-16112017-06}, we get $F_\upsilon$ \rf{man-04022017-02-a1}.

\noindent {\bf Case $h_1+1<H_1$}. This case does not respect the classical unitarity and therefore this case is not discussed in the Statement 3. This is to say that, for this case, we get classically non-unitary fields in the decomposition \rf{man-27112017-14}. Namely, from $F_\upsilon(n)$ \rf{man-27112017-12}, we see that the $|\phi^{h_2,h_1}\rangle$ describes classically unitary massive finite-component field, the $|\phi^{h_1+1,H_1}\rangle$ describes classically non-unitary partial-massless finite-component field, while the $|\phi^{H_1+1,\infty}\rangle$ describes classically unitary infinite-component spin field. Mass parameters of the fields $|\phi^{h_2,h_1}\rangle$ and $|\phi^{h_1+1,H_1}\rangle$ are given by
\beq
\label{man-27112017-30}  && m^2 = (H_1+1)^2 - h_1^2 \,, \hspace{3cm} \hbox{ for massive} \ \ |\phi^{h_2,h_1}\rangle\,,
\\
\label{man-27112017-31}  && m^2 = (h_1+1)^2 - H_1^2\,, \hspace{3cm} \hbox{ for partial-massless} \ \ |\phi^{h_1+1,H_1}\rangle\,.
\eeq
Mass parameter $m^2$ \rf{man-27112017-31} can be represented in the form
\be \label{man-27112017-32}
m^2 = - k (2H_1 - k)\,, \qquad k \equiv H_1-h_1-1\,, \hspace{1cm} \hbox{ for partial-massless} \ \ |\phi^{h_1+1,H_1}\rangle\,.
\ee
From \rf{man-27112017-32}, we learn that $|\phi^{h_1+1,H_1}\rangle$ describes depth-$k$ partial-massless mixed-symmetry field.%
\footnote{ Recent interesting discussion of mixed-symmetry partial-massless (A)dS fields may be found in Refs.\cite{Basile:2016aen,Basile:2017kaz}.}
The lowest eigenvalues of the energy operator are given by
\beq
\label{man-27112017-34}  && E_0 = H_1+3 \,, \hspace{3cm} \hbox{ for massive } \ \ |\phi^{h_2,h_1}\rangle\,,
\\
\label{man-27112017-35}  && E_0 = h_1+3\,, \hspace{3.1cm} \hbox{ for partial-massless } \ \ |\phi^{h_1+1,H_1}\rangle\,.
\eeq
The $|\phi^{h_2,h_1}\rangle$ \rf{man-27112017-14} is associated with  representation $D(E_0,h_1,h_2)$ which is unitary for $E_0$ \rf{man-27112017-34} and $h_1+1 < H_1$.
The $|\phi^{h_1+1,H_1}\rangle$ \rf{man-27112017-14} is associated with representation $D(E_0,H_1,h_2)$ which is non-unitary for $E_0$ \rf{man-27112017-35} and $h_1+1 < H_1$. Note that
$E_0$ \rf{man-27112017-35} can be represented as $E_0=H_1+2-k$, where $k$ is given in \rf{man-27112017-32}.

\noindent {\bf Irreducible totally symmetric continuous-spin field with $h_2=0$, $\kappa$-arbitrary}.  Totally symmetric fields are defined by relations \rf{man-10022018-01}, \rf{man-10022018-02}. At the end of Sec. 3, we noted that, for the case \rf{man-10022018-01}, the $\kappa$ is real-valued or purely imaginary. We consider the cases $\kappa^2>0$ and $\kappa^2 \leq 0$ in turn.

\noindent {\bf Case $h_2 = 0$, $\kappa^2>0$}. For definiteness, we assume that the $\kappa$ is strictly positive, $\kappa>0$. We note then that, for totally symmetric field, solutions to Eqs.\rf{man-26112017-24} are obtained from the {\bf Statement 1} by  setting $h_2=0$.

\noindent {\bf Case $h_2 = 0$, $\kappa^2 \leq 0$}. For this case, it is easy to see that solution to Eqs.\rf{man-26112017-24}, with $F_\upsilon$ as in \rf{man-10022018-05}, is given by
\be \label{man-11022018-01}
C_2 < \kappa^2 - 4 \,, \qquad \hbox{ for } \kappa^2 \leq 0 \,.
\ee

\noindent {\bf Reducible totally symmetric field with $h_2=0$, $\kappa$-arbitrary}. To study solutions of Eqs.\rf{man-26112017-25}, with $F_\upsilon$ as in \rf{man-10022018-05}, we consider the equation $F_\upsilon(s)=0$. Solution of this equation is given by
\be \label{man-11022018-02}
\kappa^2 = (s+1)^2 (C_2+5) - (s+1)^4\,.
\ee
Plugging $\kappa^2$ \rf{man-11022018-02} into \rf{man-10022018-05}, we get
\be \label{man-11022018-03}
F_\upsilon= \Bigl( (N_\upsilon+1)^2 - (s+1)^2\Bigr) \Bigl((N_\upsilon+1)^2 + (s+1)^2 - C_2 -5 \Bigr)\,.
\ee
On the one hand, using the notation $F_\upsilon(n) \equiv F_\upsilon|_{N_\upsilon=n}$, we get
\be \label{man-11022018-04}
F_\upsilon(n)= \Bigl( (n+1)^2 - (s+1)^2\Bigr) \Bigl((n+1)^2 + (s+1)^2 - C_2 -5 \Bigr)\,.
\ee
On the other hand, setting $h_1 = s$, $h_2=0$ in \rf{man-27112017-07}, we also obtain $F_\upsilon(n)$ as in \rf{man-11022018-04}. Taking this into account we note then that, for totally symmetric field, solutions to Eqs.\rf{man-26112017-25} are obtained from the {\bf Statements 2,3} by setting $h_1=s$, $h_2=0$.

We finish our discussion of the reducible totally symmetric field having $h_2=0$ and arbitrary $\kappa$ by the following comment. Let us decompose $\phik$ \rf{man-14112017-16} as
\be \label{man-11022018-05}
\phik = |\phi^{0,s}\rangle  + |\phi^{s+1,\infty}\rangle \,,
\ee
where $|\phi^{0,s}\rangle$, $|\phi^{s+1,\infty}\rangle$ are defined as in \rf{man-27112017-02}. If $F_\upsilon$ takes the form given in \rf{man-11022018-03}, then Lagrangian \rf{man-15112017-10} is factorized as
\be  \label{man-11022018-05-a1}
\LL  =   \LL^{0,s} + \LL^{s+1,\infty}\,,
\ee
where $\LL^{0,s}$, $\LL^{s+1,\infty}$ are defined as in \rf{man-27112017-04}, while the operator $g_\upsilon$ \rf{man-10022018-04} takes the form
\be   \label{man-11022018-06}
g_\upsilon = \Bigl( \frac{ \bigl( (N_\upsilon+1)^2 - (s+1)^2\bigr)}{ (N_\upsilon+1) (2N_\upsilon+3)  } \bigl( (N_\upsilon+1)^2 + (s+1)^2 - C_2 - 5 \bigr) \Bigr)^{1/2}\,.
\ee

\noindent {\bf Totally symmetric field with $h_2 \ne 0$, $\kappa=0$}. Now, as promised at the end of Sec. 3, we are going to demonstrate that totally symmetric field \rf{man-10022018-02} is equivalent to the infinite component field $|\phi^{s+1,\infty}\rangle$ \rf{man-11022018-05}. To prove the equivalence we should match the field contents and the operators $B^I$. First, we match the field contents.
To this end, we make the identification $h_2=s+1$, and note that $\phik$ \rf{man-14112017-16} coincides with $|\phi^{s+1,\infty}\rangle$ \rf{man-11022018-05}. Second, we match the operators $B^I$. To this end we set $\kappa=0$ in \rf{man-16112017-01}-\rf{man-16112017-06} and obtain the operator $B^I$ as in \rf{man-10022018-03},\rf{man-10022018-04} with the following expression for $g_\upsilon$
\be  \label{man-11022018-07}
g_\upsilon = \Bigl( \frac{ ((N_\upsilon+1)^2  - h_2^2 ) }{ (N_\upsilon+1) (2N_\upsilon+3)  } \bigl( (N_\upsilon+1)^2 + h_2^2 - C_2 - 5 \bigr)\Bigr)^{1/2}\,.
\ee
On the one hand, operator $B^I$ \rf{man-10022018-03},\rf{man-10022018-04}, with $g_\upsilon$ as in \rf{man-11022018-07}, describes the field $\phik$ having $h_2\ne 0$, $\kappa=0$. On the other hand, the infinite component $|\phi^{s+1,\infty}\rangle$ \rf{man-11022018-05} is described by operator $B^I$ \rf{man-10022018-03},\rf{man-10022018-04} with $g_\upsilon$ given in \rf{man-11022018-06}. Making the identification $h_2 \equiv s + 1 $, we see that the expressions for $g_\upsilon$ given in \rf{man-11022018-06} and \rf{man-11022018-07} coincide. Thus the operators $B^I$ are also matched.

\bigskip
{\bf Acknowledgments}. This work was supported by the RFBR Grant No.17-02-00546.

%%%%%%%%%%%%%%%%%%%%%%%%%%%%%%%%%%%%%%%%%%%%%%%%%%%%%%%%%%%%%%%%%%%%%%
\setcounter{section}{0} \setcounter{subsection}{0}
%%%%%%%%%%%%%%%%%%%%%%%%%%%%%%%%%%%%%%%%%%%%%%%%%%%%%%%%%%%%%%%%%%%%%%

\appendix{ \large Casimir operators of the $so(4,2)$ algebra}

To discuss Casimir operators of the $so(4,2)$ algebra it is convenient to use a manifestly 6-dim\-ensional covariant approach. In this approach, generators of the $so(4,2)$ algebra denoted by $J^{AB}$ satisfy the commutation relations
\be \label{man-28112017-01}
[J^{AB},J^{CE}]= \eta^{BC}J^{AE} + 3 \hbox{ terms},\qquad \eta^{AB} =  (--++++)\,,
\ee
where vector indices of the $so(4,2)$ algebra take values $A,B,C,E=0',0,1,2,3,4$. In terms of the generators $J^{AB}$, the Casimir operators of the $so(4,2)$ algebra can be presented as
\beq
\label{man-28112017-02} && C_2 = \half J^{A_1A_2} J^{A_2A_1}\,,
\\
\label{man-28112017-03} && C_{\epsilon\, 3} = -\frac{\irm}{48} \epsilon^{A_1\ldots A_6} J^{A_1A_2}J^{A_3A_4}J^{A_5A_6}\,,
\\
\label{man-28112017-04} && C_4 = \frac{1}{8} (J^{A_1A_2} J^{A_2A_1})^2  + \frac{3}{2} J^{A_1A_2} J^{A_2A_1}  - \frac{1}{4} J^{A_1A_2} J^{A_2A_3} J^{A_3A_4} J^{A_4A_1}\,,
\eeq
where $\epsilon^{A_1\ldots A_6}$ stands for the Levi-Civita symbol of rank six with $\epsilon^{0'01234}=1$.
Our choice of the particular form of the 4th-order Casimir operator $C_4$ \rf{man-28112017-04} is motivated by the following relation for the $C_4$ \rf{man-28112017-04}:
\beq
\label{man-28112017-05} && C_4 = \frac{1}{128} X_{\epsilon\, 2}^{AB} X_{\epsilon\, 2}^{AB}\,,
\\
\label{man-28112017-06} && X_{\epsilon\, 2}^{AB} \equiv \epsilon^{AB C_1C_2C_3C_4} J^{C_1C_2} J^{C_3C_4} \,.
\eeq
In Ref.\cite{Metsaev:1999ui}, we demonstrated that relation \rf{man-28112017-02} allows us to find the representation for the operator $A$ given in \rf{man-15112017-25}. Now, using the light-cone gauge realization of generators of the $so(4,2)$ algebra in terms of the differential operators given in  \rf{man-15112017-14}-\rf{man-15112017-22}, we can verify that the Casimir operators defined in \rf{man-28112017-03}, \rf{man-28112017-04} take the form given in \rf{man-26112017-15},\rf{man-26112017-16}. To this end we relate the 6-dimensional notation for the generators in \rf{man-28112017-01} to the 4-dimensional notation for the generators in \rf{man-27112017-36}. Namely, let us decompose 6-dimensional coordinates $x^A$ as
\be \label{man-28112017-07}
x^A = x^\oplus,\ x^\ominus\,, \ x^a\,, \qquad a=0,1,2,3\,, \qquad   x^\oplus = \frac{1}{\sqrt{2}}(x^4 +    x^{0^\prime})\,, \quad x^\ominus = \frac{1}{\sqrt{2}}(x^4  -    x^{0^\prime})\,.
\ee
It is easy then to see that, in the frame of the coordinates $x^{\oplus,\ominus}$, $x^a$,  the generators $J^{AB}$ and the flat metric tensor $\eta^{AB}$ \rf{man-28112017-01} are decomposed as
\beq
\label{man-28112017-08} && J^{AB} = J^{\oplus a}\, \  J^{\ominus a}, \ J^{\ominus \oplus},\  J^{ab}\,,
\\
\label{man-28112017-09} && \eta^{AB} = \eta^{\oplus\ominus}, \eta^{\ominus\oplus}, \eta^{ab}\,, \qquad \eta^{\oplus\ominus}=1, \quad \eta^{\ominus\oplus}=1\,.
\eeq
Generators $J^{ab}$ in \rf{man-28112017-08} are identified with the $J^{ab}$ appearing in \rf{man-27112017-36}, while the remaining generators in \rf{man-27112017-36} and \rf{man-28112017-08} are identified in the following way:
\be \label{man-28112017-10}
P^a = J^{\oplus a}\, \qquad  K^a =  J^{\ominus a}, \qquad D =  J^{\ominus \oplus}\,.
\ee
Making use of \rf{man-28112017-10} and the light-cone gauge realization for the generators in \rf{man-15112017-14}-\rf{man-15112017-22}, we verified that expressions for $C_{\epsilon\, 3}$, $C_4$ in \rf{man-28112017-03},\rf{man-28112017-04} lead to expressions for $C_{\epsilon\, 3}$, $C_4$ in \rf{man-26112017-15},\rf{man-26112017-16}.

For the $D(E_0,h_1,h_2)$, which is positive-energy lowest weight representation of the $so(4,2)$ algebra, eigenvalues of the Casimir operators \rf{man-28112017-02}-\rf{man-28112017-04} are given by
\beq
&& C_2 = E_0(E_0-4) + h_1(h_1+2) + h_2^2\,,
\nonumber\\
\label{man-28112017-12} && C_{\epsilon\, 3} = (E_0-2)(h_1+1)h_2\,,
\\
&& C_4 = (E_0-1)(E_0-3) h_1(h_1+2) + h_2^2 \bigl(E_0(E_0-4) + h_1 (h_1+2) + 4 \bigr)\,.
\nonumber
\eeq

\appendix{ \large Useful relations for various spin operators }

Creation operators $\alpha^I$ and the respective annihilation operators $\alphab^I$ are referred to as oscillators. The oscillators, hermitian conjugation rule, and the vacuum $|0\rangle$ are defined by the relations
\be \label{man-12022018-01}
[\alphab^I,\alpha^J] = \delta^{IJ}\,, \qquad \alpha^{I\dagger} = \alphab^I\,, \qquad  \alphab^I|0\rangle\,,
\ee
where $\delta^{IJ}$ stands for the Kronecker delta symbol. Vector indices of the $so(3)$ algebra take values $I,J,K=1,2,3$. We use the following notation for various operators constructed out of the oscillators
\beq
\label{man-12022018-02} && S^I \equiv \epsilon^{IJK}\alpha^J \alphab^K \,,
\\
\label{man-12022018-03} &&  M^{IJ} \equiv \alpha^I \alphab^J - \alpha^J \alphab^I\,,
\\
\label{man-12022018-04} && A^I \equiv \alpha^I -  \alpha^2 \frac{1}{2N_\alpha+3} \alphab^I\,,
\\
\label{man-12022018-05} && N_\alpha \equiv \alpha^I\alphab^I\,, \qquad \alpha^2 \equiv \alpha^I\alpha^I\,, \qquad \alphab^2 \equiv \alphab^I\alphab^I\,,
\eeq
where $\epsilon^{IJK}$ stands for the Levi-Civita symbol of rank three with $\epsilon^{123}=1$. The operators $M^{IJ}$ and $S^I$ are related as
\be
M^{IJ} = \epsilon^{IJK} S^K\,, \qquad  S^I = \half  \epsilon^{IJK} M^{JK}\,.
\ee
For $M^{IJ}$ and $S^I$, we note the following commutation relations and hermitian conjugation rules
\beq
&& [M^{IJ},M^{KL}] = \delta^{JK} M^{IL}+3\hbox{ terms}\,, \qquad M^{IJ\dagger} = - M^{IJ}\,,
\\
&& [S^I,S^J]= - M^{IJ}\,, \hspace{3.8cm} S^{I\dagger} = - S^I.
\eeq
For studying the defining equations in \rf{man-14112017-14}, \rf{man-15112017-28},
the following relations turn out to be helpful
\beq
&& A^I A^J  -A^J A^I =0\,,
\\
&& \alphab^I S^J  - \alphab^J S^I = - \epsilon^{IJK} (N_\alpha +2) \alphab^K + \epsilon^{IJK} \alpha^K \bar\alpha^2\,,
\\
&& S^I  \alphab^J - S^J \alphab^I = \epsilon^{IJK} N_\alpha \alphab^K - \epsilon^{IJK} \alpha^K \alphab^2\,,
\\
&& A^I S^J - A^J S^I = \epsilon^{IJK} A^K N_\alpha - \epsilon^{IJK} \alpha^K \alpha^2 \frac{1}{2N_\alpha+5}\alphab^2\,,
\\
&&  S^I A^J  -  S^J A^I = -\epsilon^{IJK}  A^K (N_\alpha+2)  + \epsilon^{IJK}
\alpha^K \alpha^2 \frac{1}{2N_\alpha+5} \alphab^2\,,
\\[5pt]
&& A^I \alphab^J - A^J \alphab^I = M^{IJ}\,,
\\
&& \alphab^I A^J - \alphab^J A^I  = - \frac{2N_\alpha+3}{2N_\alpha+1} M^{IJ}\,,
\\
&& M^{IJ} A^J = A^I (N_\alpha+2) - \alpha^I\alpha^2 \frac{1}{2N_\alpha+5} \alphab^2\,,
\\
&& M^{IJ} \alphab^J  = -N_\alpha \alphab^I + \alpha^I \alphab^2\,,
\\
&& M^{IJ} S^J = S^I\,,
\\
&& M^{IL}M^{LK}M^{KJ} - (I\leftrightarrow J)= -  M^{KL}M^{KL} M^{IJ}\,.
\eeq

For the computation of eigenvalues of the Casimir operators \rf{man-26112017-05},\rf{man-26112017-06} and \rf{man-26112017-15},\rf{man-26112017-16}, we use the following relations:
\beq
&& A^I \alphab^I  = N_\alpha - \alpha^2 \frac{1}{2N_\alpha+3}\alphab^2\,,
\\
&& \alphab^I A^I = \frac{(2N_\alpha + 3)(N_\alpha+1)}{2N_\alpha + 1} -  \alpha^2 \frac{1}{2N_\alpha+5}\alphab^2\,,
\\
&& S^I S^I =  - N_\alpha(N_\alpha+1) + \alpha^2\alphab^2 \,,
\\
&& M^{IJ}M^{IJ} = 2S^IS^I\,.
\eeq


\small

\begin{thebibliography}{30}

\parskip-5pt



%\cite{Bekaert:2005in}
\bibitem{Bekaert:2005in}
  X.~Bekaert and J.~Mourad,
  %``The Continuous spin limit of higher spin field equations,''
  JHEP {\bf 0601}, 115 (2006)
  %doi:10.1088/1126-6708/2006/01/115
  [hep-th/0509092].
  %%CITATION = doi:10.1088/1126-6708/2006/01/115;%%


%\cite{Bekaert:2006py}
\bibitem{Bekaert:2006py}
  X.~Bekaert and N.~Boulanger,
 ``The Unitary representations of the Poincare group in any spacetime dimension,''
   in 2nd Modave Summer School in Theoretical Physics Modave, Belgium, August 6-12, 2006, 2006.
  hep-th/0611263.
  %%CITATION = HEP-TH/0611263;%%


%\cite{Brink:2002zx}
\bibitem{Brink:2002zx}
  L.~Brink, A.~M.~Khan, P.~Ramond and X.~z.~Xiong,
  %``Continuous spin representations of the Poincare and superPoincare groups,''
  J.\ Math.\ Phys.\  {\bf 43}, 6279 (2002)
  %doi:10.1063/1.1518138
  [hep-th/0205145].
  %%CITATION = doi:10.1063/1.1518138;%%



%\cite{Bengtsson:2013vra}
\bibitem{Bengtsson:2013vra}
  A.~K.~H.~Bengtsson,
  %``BRST Theory for Continuous Spin,''
  JHEP {\bf 1310}, 108 (2013)
%  doi:10.1007/JHEP10(2013)108
  [arXiv:1303.3799 [hep-th]].
  %%CITATION = doi:10.1007/JHEP10(2013)108;%%


%\cite{Schuster:2014hca}
\bibitem{Schuster:2014hca}
  P.~Schuster and N.~Toro,
  %``Continuous-spin particle field theory with helicity correspondence,''
  Phys.\ Rev.\ D {\bf 91}, 025023 (2015)
  %doi:10.1103/PhysRevD.91.025023
  [arXiv:1404.0675 [hep-th]].
  %%CITATION = doi:10.1103/PhysRevD.91.025023;%%



%\cite{Najafizadeh:2015uxa}
\bibitem{Najafizadeh:2015uxa}
  X.Bekaert, M.Najafizadeh, M.R.Setare,
  %``A gauge field theory of fermionic Continuous-Spin Particles,''
  Phys.\ Lett.\ B {\bf 760}, 320 (2016)
% doi:10.1016/j.physletb.2016.07.005
  [arXiv:1506.00973 [hep-th]].
  %%CITATION = doi:10.1016/j.physletb.2016.07.005;%%



%\cite{Rivelles:2014fsa}
\bibitem{Rivelles:2014fsa}
  V.~O.~Rivelles,
  %``Gauge Theory Formulations for Continuous and Higher Spin Fields,''
  Phys.\ Rev.\ D {\bf 91}, no. 12, 125035 (2015)
% doi:10.1103/PhysRevD.91.125035
  [arXiv:1408.3576 [hep-th]].
  %%CITATION = doi:10.1103/PhysRevD.91.125035;%%




%\cite{Metsaev:2016lhs}
\bibitem{Metsaev:2016lhs}
  R.~R.~Metsaev,
  %``Continuous spin gauge field in (A)dS space,''
  Phys.\ Lett.\ B {\bf 767}, 458 (2017)
 % doi:10.1016/j.physletb.2017.02.027
  [arXiv:1610.00657 [hep-th]].


%\cite{Metsaev:2017ytk}
\bibitem{Metsaev:2017ytk}
  R.~R.~Metsaev,
  %``Fermionic continuous spin gauge field in (A)dS space,''
  Phys.\ Lett.\ B {\bf 773}, 135 (2017)
%  doi:10.1016/j.physletb.2017.08.020
  [arXiv:1703.05780 [hep-th]].
  %%CITATION = doi:10.1016/j.physletb.2017.08.020;%%


%\cite{Savvidy:2003fx}
\bibitem{Savvidy:2003fx}
  G.~K.~Savvidy,
  %``Tensionless strings: Physical Fock space and higher spin fields,''
  Int.\ J.\ Mod.\ Phys.\ A {\bf 19}, 3171 (2004)
  %doi:10.1142/S0217751X04018312
  [hep-th/0310085].
  %%CITATION = doi:10.1142/S0217751X04018312;%%
%
\\[-2pt]
%
%\cite{Mourad:2005rt}
%\bibitem{Mourad:2005rt}
  J.~Mourad,
  ``Continuous spin particles from a string theory,''
  hep-th/0504118.
  %%CITATION = HEP-TH/0504118;%%



%\cite{Zinoviev:2017rnj}
\bibitem{Zinoviev:2017rnj}
  Y.~M.~Zinoviev,
  %``Infinite spin fields in d = 3 and beyond,''
  Universe {\bf 3}, no. 3, 63 (2017)
%  doi:10.3390/universe3030063
  [arXiv:1707.08832 [hep-th]].
  %%CITATION = doi:10.3390/universe3030063;%%


%\cite{Najafizadeh:2017tin}
\bibitem{Najafizadeh:2017tin} 
  M.~Najafizadeh,
  %``Modified Wigner equations and continuous spin gauge field,''
  Phys.\ Rev.\ D {\bf 97}, no. 6, 065009 (2018)
%  doi:10.1103/PhysRevD.97.065009
  [arXiv:1708.00827 [hep-th]].
  %%CITATION = doi:10.1103/PhysRevD.97.065009;%%


%\cite{Bekaert:2017khg}
\bibitem{Bekaert:2017khg}
  X.~Bekaert and E.~D.~Skvortsov,
  %``Elementary particles with continuous-spin,''
  Int.J.Mod.Phys.\ A {\bf 32}, no.23n24, 1730019 (2017)
% doi:10.1142/S0217751X17300198
  [arXiv:1708.01030].
  %%CITATION = doi:10.1142/S0217751X17300198;%%


%\cite{Metsaev:2017cuz}
\bibitem{Metsaev:2017cuz}
  R.~R.~Metsaev,
  %``Cubic interaction vertices for continuous-spin fields and arbitrary spin massive fields,''
  JHEP {\bf 1711}, 197 (2017)
%  doi:10.1007/JHEP11(2017)197
  [arXiv:1709.08596 [hep-th]].
  %%CITATION = doi:10.1007/JHEP11(2017)197;%%

%\cite{Bekaert:2017xin}
\bibitem{Bekaert:2017xin}
  X.~Bekaert, J.~Mourad and M.~Najafizadeh,
  %``Continuous-spin field propagator and interaction with matter,''
  JHEP {\bf 1711}, 113 (2017)
%  doi:10.1007/JHEP11(2017)113
  [arXiv:1710.05788 [hep-th]].
  %%CITATION = doi:10.1007/JHEP11(2017)113;%%
  %1 citations counted in INSPIRE as of 28 Nov 2017



%\cite{Khabarov:2017lth}
\bibitem{Khabarov:2017lth}
  M.~V.~Khabarov and Y.~M.~Zinoviev,
  %``Infinite (continuous) spin fields in the frame-like formalism,''
  Nucl.\ Phys.\ B {\bf 928}, 182 (2018)
%  doi:10.1016/j.nuclphysb.2018.01.016
  [arXiv:1711.08223 [hep-th]].
  %%CITATION = doi:10.1016/j.nuclphysb.2018.01.016;%%



%\cite{Vasiliev:1990en}
\bibitem{Vasiliev:1990en}
  M.~A.~Vasiliev,
  %``Consistent equation for interacting gauge fields of all spins in (3+1)-dimensions,''
  Phys.\ Lett.\ B {\bf 243}, 378 (1990).
%  doi:10.1016/0370-2693(90)91400-6
  %%CITATION = doi:10.1016/0370-2693(90)91400-6;%%
%
\\[-2pt]
%
%\cite{Vasiliev:2003ev}
%\bibitem{Vasiliev:2003ev}
  M.~A.~Vasiliev,
  %``Nonlinear equations for symmetric massless higher spin fields in
  %(A)dS(d),''
  Phys.\ Lett.\  B {\bf 567}, 139 (2003)
  [arXiv:hep-th/0304049].
  %%CITATION = PHLTA,B567,139;%%



%\cite{Metsaev:1998it}
\bibitem{Metsaev:1998it}
  R.~R.~Metsaev and A.~A.~Tseytlin,
  %``Type IIB superstring action in AdS(5) x S**5 background,''
  Nucl.\ Phys.\ B {\bf 533}, 109 (1998)
%  doi:10.1016/S0550-3213(98)00570-7
  [hep-th/9805028].
  %%CITATION = doi:10.1016/S0550-3213(98)00570-7;%%


%\cite{Metsaev:2000yf}
\bibitem{Metsaev:2000yf}
  R.~R.~Metsaev and A.~A.~Tseytlin,
  %``Superstring action in AdS(5) x S**5. Kappa symmetry light cone gauge,''
  Phys.\ Rev.\ D {\bf 63}, 046002 (2001)
%  doi:10.1103/PhysRevD.63.046002
  [hep-th/0007036].
  %%CITATION = doi:10.1103/PhysRevD.63.046002;%%


%\cite{Metsaev:2000yu}
\bibitem{Metsaev:2000yu}
  R.~R.~Metsaev, C.~B.~Thorn and A.~A.~Tseytlin,
  %``Light cone superstring in AdS space-time,''
  Nucl.\ Phys.\ B {\bf 596}, 151 (2001)
%  doi:10.1016/S0550-3213(00)00712-4
  [hep-th/0009171].
  %%CITATION = doi:10.1016/S0550-3213(00)00712-4;%%

%\cite{Metsaev:1999ui}
\bibitem{Metsaev:1999ui}
  R.~R.~Metsaev,
  %``Light cone form of field dynamics in Anti-de Sitter space-time and AdS / CFT correspondence,''
  Nucl.\ Phys.\ B {\bf 563}, 295 (1999)
%  doi:10.1016/S0550-3213(99)00554-4
  [hep-th/9906217].
  %%CITATION = doi:10.1016/S0550-3213(99)00554-4;%%



%\cite{Metsaev:2002vr}
\bibitem{Metsaev:2002vr}
  R.~R.~Metsaev,
  %``Massless arbitrary spin fields in AdS(5),''
  Phys.\ Lett.\ B {\bf 531}, 152 (2002)
%  doi:10.1016/S0370-2693(02)01344-8
  [hep-th/0201226].
  %%CITATION = doi:10.1016/S0370-2693(02)01344-8;%%


%\cite{Metsaev:2003cu}
\bibitem{Metsaev:2003cu}
R.~R.~Metsaev,
%``Massive totally symmetric fields in AdS(d),''
Phys.\ Lett.\ B {\bf 590}, 95 (2004) [arXiv:hep-th/0312297].
%%CITATION = HEP-TH 0312297;%%


%\cite{Metsaev:2004ee}
\bibitem{Metsaev:2004ee}
  R.~R.~Metsaev,
  %``Mixed symmetry massive fields in AdS(5),''
  Class.\ Quant.\ Grav.\  {\bf 22}, 2777 (2005)
%  doi:10.1088/0264-9381/22/13/016
  [hep-th/0412311].
  %%CITATION = doi:10.1088/0264-9381/22/13/016;%%


%%%%%%%%%%%%%%%%%%%%%%%%%%%%%%%%%%%%%%%%%%%%%%%%%%%%%%%%%%%%%%%%%%%%%%%%%%%%%%%%%%%%%%%%%%%
%%%%%%%%%%%%%%%%%%%%%%%%%%%%%%%%%%%%%%%%%%%%%%%%%%%%%%%%%%%%%%%%%%%%%%%%%%%%%%%%%%%%%%%%%%%


%\cite{Alkalaev:2005kw}
\bibitem{Alkalaev:2005kw}
  K.~B.~Alkalaev, O.~V.~Shaynkman and M.~A.~Vasiliev,
  %``Lagrangian formulation for free mixed-symmetry bosonic gauge fields in (A)dS(d),''
  JHEP {\bf 0508}, 069 (2005)
  [hep-th/0501108].
  %%CITATION = HEP-TH/0501108;%%


%\cite{Campoleoni:2008jq}
\bibitem{Campoleoni:2008jq}
  A.~Campoleoni, D.~Francia, J.~Mourad, A.~Sagnotti,
  %``Unconstrained Higher Spins of Mixed Symmetry. I. Bose Fields,''
  Nucl.Phys. B {\bf 815}, 289 (2009)
  [arXiv:0810.4350]
  %%CITATION = ARXIV:0810.4350;%%


%\cite{Boulanger:2008up}
\bibitem{Boulanger:2008up}
  N.~Boulanger, C.~Iazeolla and P.~Sundell,
  %``Unfolding Mixed-Symmetry Fields in AdS and the BMV Conjecture: I. General Formalism,''
  JHEP {\bf 0907}, 013 (2009)
  [arXiv:0812.3615 [hep-th]].
  %%CITATION = ARXIV:0812.3615;%%
%
\\
%
%\cite{Boulanger:2008kw}
%\bibitem{Boulanger:2008kw}
  N.~Boulanger, C.~Iazeolla and P.~Sundell,
  %``Unfolding Mixed-Symmetry Fields in AdS and the BMV Conjecture. II. Oscillator Realization,''
  JHEP {\bf 0907}, 014 (2009)
  [arXiv:0812.4438 [hep-th]].
  %%CITATION = ARXIV:0812.4438;%%
%
\\
%
%\cite{Alkalaev:2009vm}
%\bibitem{Alkalaev:2009vm}
  K.~B.~Alkalaev and M.~Grigoriev,
  %``Unified BRST description of AdS gauge fields,''
  Nucl.\ Phys.\ B {\bf 835} (2010) 197
%  doi:10.1016/j.nuclphysb.2010.04.004
  [arXiv:0910.2690 [hep-th]].
  %%CITATION = doi:10.1016/j.nuclphysb.2010.04.004;%%
%
\\
%
%\cite{Chekmenev:2015kzf}
%\bibitem{Chekmenev:2015kzf}
  A.~Chekmenev and M.~Grigoriev,
  %``Boundary values of mixed-symmetry massless fields in AdS space,''
  Nucl.\ Phys.\ B {\bf 913}, 769 (2016)
%  doi:10.1016/j.nuclphysb.2016.10.006
  [arXiv:1512.06443 [hep-th]].
  %%CITATION = doi:10.1016/j.nuclphysb.2016.10.006;%%


%\cite{Fotopoulos:2010ay}
\bibitem{Fotopoulos:2010ay}
  A.~Fotopoulos and M.~Tsulaia,
  %``On the Tensionless Limit of String theory,
  %Off - Shell Higher Spin Interaction Vertices and BCFW Recursion Relations,''
  JHEP {\bf 1011}, 086 (2010)
  [arXiv:1009.0727 [hep-th]].
  %%CITATION = ARXIV:1009.0727;%%
%
\\
%
%\cite{Buchbinder:2011xw}
%\bibitem{Buchbinder:2011xw}
  I.~L.~Buchbinder and A.~Reshetnyak,
  %``General Lagrangian Formulation for Higher Spin Fields with Arbitrary Index Symmetry. I. Bosonic fields,''
  Nucl.\ Phys.\ B {\bf 862}, 270 (2012)
%  doi:10.1016/j.nuclphysb.2012.04.016
  [arXiv:1110.5044 [hep-th]].
  %%CITATION = doi:10.1016/j.nuclphysb.2012.04.016;%%
%
\\
%
%\cite{Reshetnyak:2010ga}
%\bibitem{Reshetnyak:2010ga}
  A.~A.~Reshetnyak,
  %``Towards Lagrangian formulations of mixed-symmetry Higher Spin Fields on AdS-space within BFV-BRST formalism,''
  Phys.\ Part.\ Nucl.\  {\bf 41}, 976 (2010)
  [arXiv:1002.0124 [hep-th]].


%\cite{Zinoviev:2009vy}
\bibitem{Zinoviev:2009vy}
  Y.~M.~Zinoviev,
  %``Frame-like gauge invariant formulation for mixed symmetry fermionic fields,''
  Nucl.\ Phys.\ B {\bf 821}, 21 (2009)
%  doi:10.1016/j.nuclphysb.2009.06.008
  [arXiv:0904.0549 [hep-th]].
  %%CITATION = doi:10.1016/j.nuclphysb.2009.06.008;%%
%
\\
%
%\cite{Zinoviev:2009gh}
%\bibitem{Zinoviev:2009gh}
  Y.~M.~Zinoviev,
  %``Towards frame-like gauge invariant formulation for massive mixed symmetry bosonic fields. II. General Young tableau with two rows,''
  Nucl.\ Phys.\ B {\bf 826}, 490 (2010)
%  doi:10.1016/j.nuclphysb.2009.08.019
  [arXiv:0907.2140 [hep-th]].
  %%CITATION = doi:10.1016/j.nuclphysb.2009.08.019;%%
%
\\
%
%\cite{Skvortsov:2010nh}
%\bibitem{Skvortsov:2010nh}
  E.~D.~Skvortsov and Y.~M.~Zinoviev,
  %``Frame-like Actions for Massless Mixed-Symmetry Fields in Minkowski space. Fermions,''
  Nucl.\ Phys.\ B {\bf 843}, 559 (2011)
  %doi:10.1016/j.nuclphysb.2010.10.012
  [arXiv:1007.4944 [hep-th]].
  %%CITATION = doi:10.1016/j.nuclphysb.2010.10.012;%%



%\cite{Alkalaev:2006hq}
\bibitem{Alkalaev:2006hq}
  K.~B.~Alkalaev,
  %``Mixed-symmetry massless gauge fields in AdS(5),''
  Theor.\ Math.\ Phys.\  {\bf 149}, 1338 (2006)
%  [Teor.\ Mat.\ Fiz.\  {\bf 149}, 47 (2006)]
%  doi:10.1007/s11232-006-0122-5
  [hep-th/0501105].
%
\\
%\cite{Sorokin:2017irs}
%\bibitem{Sorokin:2017irs}
  D.~Sorokin and M.~Tsulaia,
%  ``Higher Spin Fields in Hyperspace. A Review,''
  Universe {\bf 4}, no. 1, 7 (2018)
%  doi:10.3390/universe4010007
  [arXiv:1710.08244 [hep-th]].
  %%CITATION = doi:10.3390/universe4010007;%%%
\\
%
%\cite{Skvortsov:2016lbh}
%\bibitem{Skvortsov:2016lbh}
  E.~Skvortsov, D.~Sorokin and M.~Tsulaia,
  %``Correlation Functions of Sp(2n) Invariant Higher-Spin Systems,''
  JHEP {\bf 1607}, 128 (2016)
% doi:10.1007/JHEP07(2016)128
  [arXiv:1605.08498 [hep-th]].
  %%CITATION = doi:10.1007/JHEP07(2016)128;%%
%
\\
%
%\cite{Adamo:2016ple}
%\bibitem{Adamo:2016ple}
  T.~Adamo, P.~Hähnel and T.~McLoughlin,
  %``Conformal higher spin scattering amplitudes from twistor space,''
  JHEP {\bf 1704}, 021 (2017)
% doi:10.1007/JHEP04(2017)021
  [arXiv:1611.06200 [hep-th]].
  %%CITATION = doi:10.1007/JHEP04(2017)021;%%
%
\\
%
%\cite{Haehnel:2016mlb}
%\bibitem{Haehnel:2016mlb}
  P.~Haehnel and T.~McLoughlin,
  %``Conformal Higher Spin Theory and Twistor Space Actions,''
  arXiv:1604.08209 [hep-th].
  %%CITATION = ARXIV:1604.08209;%%
%
\\
%
%\cite{Uvarov:2016slb}
%\bibitem{Uvarov:2016slb}
  D.~V.~Uvarov,
  %``Ambitwistors, oscillators and massless fields on $AdS_5$,''
  Phys.\ Lett.\ B {\bf 762}, 415 (2016)
% doi:10.1016/j.physletb.2016.09.065
  [arXiv:1607.05233 [hep-th]].
%
\\
%
%\cite{Uvarov:2017lyu}
%\bibitem{Uvarov:2017lyu}
  D.V.~Uvarov,
  Spinning particle and null-string on $AdS_d$: projective-space approach,''
  [1707.05761]



%%%%%%%%%%%%%%%%%%%%%%%%%%%%%%%%%%%%%%%%%%%%%%%%%%%%%%%%%%%%%%%%%%%%%%%%%%%%%%%%%%%%%%%%%%%
%%%%%%%%%%%%%%%%%%%%%%%%%%%%%%%%%%%%%%%%%%%%%%%%%%%%%%%%%%%%%%%%%%%%%%%%%%%%%%%%%%%%%%%%%%%


%\cite{Metsaev:2005ar}
\bibitem{Metsaev:2005ar}
  R.~R.~Metsaev,
  %``Cubic interaction vertices of massive and massless higher spin fields,''
  Nucl.\ Phys.\ B {\bf 759}, 147 (2006)
%  doi:10.1016/j.nuclphysb.2006.10.002
  [hep-th/0512342].
  %%CITATION = doi:10.1016/j.nuclphysb.2006.10.002;%%

%\cite{Metsaev:2007rn}
\bibitem{Metsaev:2007rn}
  R.~R.~Metsaev,
  %``Cubic interaction vertices for fermionic and bosonic arbitrary spin fields,''
  Nucl.\ Phys.\ B {\bf 859}, 13 (2012)
 % doi:10.1016/j.nuclphysb.2012.01.022
  [arXiv:0712.3526 [hep-th]].
  %%CITATION = doi:10.1016/j.nuclphysb.2012.01.022;%%


%\cite{Ponomarev:2016lrm}
\bibitem{Ponomarev:2016lrm}
  D.~Ponomarev and E.~D.~Skvortsov,
  %``Light-Front Higher-Spin Theories in Flat Space,''
  J.\ Phys.\ A {\bf 50}, no. 9, 095401 (2017)
  %doi:10.1088/1751-8121/aa56e7
  [arXiv:1609.04655 [hep-th]].
  %%CITATION = doi:10.1088/1751-8121/aa56e7;%%
%
\\[-2pt]
%
%\cite{Sleight:2016xqq}
%\bibitem{Sleight:2016xqq}
  C.~Sleight and M.~Taronna,
  %``Higher-Spin Algebras, Holography and Flat Space,''
  JHEP {\bf 1702}, 095 (2017)
  %doi:10.1007/JHEP02(2017)095
  [arXiv:1609.00991 [hep-th]].
  %%CITATION = doi:10.1007/JHEP02(2017)095;%%
%
\\
%
%\cite{Conde:2016izb}
%\bibitem{Conde:2016izb}
  E.~Conde, E.~Joung and K.~Mkrtchyan,
  %``Spinor-Helicity Three-Point Amplitudes from Local Cubic Interactions,''
  JHEP {\bf 1608}, 040 (2016)
%  doi:10.1007/JHEP08(2016)040
  [arXiv:1605.07402 [hep-th]].
  %%CITATION = doi:10.1007/JHEP08(2016)040;%%
%
\\
%
%\cite{Ananth:2016abv}
%\bibitem{Ananth:2016abv}
  S.~Ananth, L.~Brink and S.~Majumdar,
  %``Exceptional versus superPoincaré algebra as the defining symmetry of maximal supergravity,''
  JHEP {\bf 1603}, 051 (2016)
%  doi:10.1007/JHEP03(2016)051
  [arXiv:1601.02836 [hep-th]].
  %%CITATION = doi:10.1007/JHEP03(2016)051;%%
%
\\
%
%\cite{Ananth:2017xpj}
%\bibitem{Ananth:2017xpj}
  S.~Ananth, L.~Brink, S.~Majumdar, M.~Mali and N.~Shah,
  %``Gravitation and quadratic forms,''
  JHEP {\bf 1703}, 169 (2017)
%  doi:10.1007/JHEP03(2017)169
  [arXiv:1702.06261].
  %%CITATION = doi:10.1007/JHEP03(2017)169;%%
%
\\
%
%\cite{Ponomarev:2017nrr}
%\bibitem{Ponomarev:2017nrr} 
  D.~Ponomarev,
  %``Chiral Higher Spin Theories and Self-Duality,''
  JHEP {\bf 1712}, 141 (2017)
%  doi:10.1007/JHEP12(2017)141
  [arXiv:1710.00270 [hep-th]].
  %%CITATION = doi:10.1007/JHEP12(2017)141;%%



%\cite{Metsaev:2016rpa}
\bibitem{Metsaev:2016rpa}
  R.~R.~Metsaev,
  ``Interacting light-cone gauge conformal fields,''
  arXiv:1612.06348 [hep-th].
  %%CITATION = ARXIV:1612.06348;%%

  %\cite{Metsaev:2014sfa}
\bibitem{Metsaev:2014sfa}
  R.~R.~Metsaev,
  %``Mixed-symmetry fields in AdS(5), conformal fields, and AdS/CFT,''
  JHEP {\bf 1501}, 077 (2015)
%  doi:10.1007/JHEP01(2015)077
  [arXiv:1410.7314 [hep-th]].
  %%CITATION = doi:10.1007/JHEP01(2015)077;%%

%\cite{Metsaev:2015rda}
\bibitem{Metsaev:2015rda}
  R.~R.~Metsaev,
  %``Light-cone AdS/CFT-adapted approach to AdS fields/currents, shadows, and conformal fields,''
  JHEP {\bf 1510}, 110 (2015)
%  doi:10.1007/JHEP10(2015)110
  [arXiv:1507.06584 [hep-th]].
  %%CITATION = doi:10.1007/JHEP10(2015)110;%%




%\cite{Basile:2016aen}
\bibitem{Basile:2016aen}
  T.~Basile, X.~Bekaert and N.~Boulanger,
  %``Mixed-symmetry fields in de Sitter space: a group theoretical glance,''
  JHEP {\bf 1705}, 081 (2017)
% doi:10.1007/JHEP05(2017)081
  [arXiv:1612.08166 [hep-th]].
  %%CITATION = doi:10.1007/JHEP05(2017)081;%%

%\cite{Basile:2017kaz}
\bibitem{Basile:2017kaz}
  T.~Basile,
  %``A note on rectangular partially massless fields,''
  Universe {\bf 4}, no. 1, 4 (2018)
%  doi:10.3390/universe4010004
  [arXiv:1710.10572 [hep-th]].
  %%CITATION = doi:10.3390/universe4010004;%%






\end{thebibliography}
\end{document}